\documentclass{pasj01}
\Received{19 August 2022}
\Accepted{21 November 2022}
\Published{$\langle$publication date$\rangle$}
\usepackage{scalefnt}
\usepackage{natbib}
\usepackage{url}
\usepackage{ulem}
\usepackage{wasysym} 
\usepackage{bm} 
\usepackage{ mathrsfs } 

\begin{document}

\title{EAVN Astrometry toward the Extreme Outer Galaxy: Kinematic distance with the proper motion of G034.84$-$00.95}
\author{
Nobuyuki \textsc{Sakai}\altaffilmark{1, 2}, Bo Zhang\altaffilmark{3}, Shuangjing Xu\altaffilmark{1, 3}, 
Daisuke \textsc{Sakai}\altaffilmark{4, 5}, Yoshiaki \textsc{Tamura}\altaffilmark{4}, Takaaki \textsc{Jike}\altaffilmark{4}, Taehyun Jung\altaffilmark{1}, Chungsik Oh\altaffilmark{1}, Jeong-Sook Kim\altaffilmark{6}, Noriyuki \textsc{Kawaguchi}\altaffilmark{7}, Hiroshi \textsc{Imai}\altaffilmark{8}, Wu \textsc{Jiang}\altaffilmark{3}, Lang \textsc{Cui}\altaffilmark{9}, Soon-Wook Kim\altaffilmark{1}, Pengfei \textsc{Jiang}\altaffilmark{9}, Tomoharu Kurayama\altaffilmark{10}, Jeong Ae Lee\altaffilmark{1}, Kazuya Hachisuka\altaffilmark{4}, Dong-Kyu Jung\altaffilmark{1}, Bo Xia\altaffilmark{3}, Guanghui Li\altaffilmark{9}, Mareki Honma\altaffilmark{4}, Kee-Tae Kim\altaffilmark{1}, Zhi-Qiang Shen\altaffilmark{3, 11}, and Na Wang\altaffilmark{9, 11}.
}%
\altaffiltext{1}{Korea Astronomy $\&$ Space Science Institute, 776, Daedeokdae-ro, Yuseong-gu, Daejeon 34055, Korea}

\altaffiltext{2}{National Astronomical Research Institute of Thailand (Public Organization), 260 Moo 4, T. Donkaew, A. Maerim, Chiang Mai, 50180, Thailand}

\altaffiltext{3}{Shanghai Astronomical Observatory, Chinese Academy of Sciences,  80 Nandan Road, Shanghai, 200030, China}

\altaffiltext{4}{Mizusawa VLBI Observatory, National Astronomical
  Observatory of Japan, 2-12 Hoshigaoka, Mizusawa, Oshu, Iwate 023-0861, Japan}

\altaffiltext{5}{The Iwate Nippo Co., Ltd., 3-7 Uchimaru, Morioka, Iwate 020-8622, Japan}

\altaffiltext{6}{Ulsan National Institute of Science and Technology
UNIST-gil 50, Eonyang-eup, Ulju-gun, Ulsan 44919, Repblic of Korea}

\altaffiltext{7}{Mizusawa VLBI Observatory, National Astronomical
 Observatory of Japan, 2-21-1 Osawa, Mitaka, Tokyo 181-8588, Japan}

 \altaffiltext{8}{Amanogawa Galaxy Astronomy Research Center, Graduate School of Science and Engineering, Kagoshima University, 1-21-35 Korimoto Kagoshima}
 
 \altaffiltext{9}{Xinjiang Astronomical Observatory, Chinese Academy of Sciences, Urumqi 830011, China}
 
 \altaffiltext{10}{Faculty of Education and Human Sciences, Department of School Education, Teikyo University of Science, 2-2-1 Senju-Sakuragi, Adachi, Tokyo 120-0045, Japan}

\altaffiltext{11}{Key Laboratory of Radio Astronomy, Chinese Academy of Sciences, Nanjing 210008, China}

 \email{nobuyuki@narit.or.th}


\maketitle

\begin{abstract}
We aim to reveal the structure and kinematics of the Outer-Scutum-Centaurus (OSC) arm located on the far side of the Milky Way through very long baseline interferometry (VLBI) astrometry using KaVA, which is composed of KVN (Korean VLBI Network) and VERA (VLBI Exploration of Radio Astrometry). We report the proper motion of a 22 GHz H$_{2}$O maser source, which is associated with the star-forming region G034.84$-$00.95, to be ($\mu_{\alpha} \rm{cos}\delta$, $\mu_{\delta}$) = ($-$1.61$\pm$0.18, $-$4.29$\pm$0.16) mas yr$^{-1}$ in equatorial coordinates (J2000). We estimate the 2D kinematic distance to the source to be 18.6$\pm$1.0 kpc, which is derived from the variance-weighted average of kinematic distances with LSR velocity and the Galactic-longitude component of the measured proper motion. Our result places the source in the OSC arm and implies that G034.84$-$00.95 is moving away from the Galactic plane with a vertical velocity of $-$38$\pm$16 km s$^{-1}$. Since the H {\scriptsize I} supershell GS033+06$-$49 is located at a kinematic distance roughly equal to that of G034.84$-$00.95, it is expected that gas circulation occurs between the outer Galactic disk around G034.84$-$00.95 with a Galactocentric distance of 12.8$^{+1.0}_{-0.9}$ kpc and halo. We evaluate possible origins of the fast vertical motion of G034.84$-$00.95, which are (1) supernova explosions and (2) cloud collisions with the Galactic disk. However, neither of the possibilities are matched with the results of VLBI astrometry as well as spatial distributions of H {\scriptsize II} regions and H {\scriptsize I} gas.


\end{abstract}

\section{Introduction}
$Gaia$ satellite and VLBI arrays have measured trigonometric parallaxes with accuracies of 10 microarcsecond ($\mu$as) or better for stars and masers associated with star-forming regions, respectively (e.g. \citealp{2022yCat.1355....0G}; \citealp{2019ApJ...885..131R}; \citealp{2020PASJ...72...50V}). This allows us to make a 3D map of the Milky Way. However, the current parallax measurements are conducted mostly within 10 kpc from the Sun, which delineates roughly one-third of the Galactic disk (e.g., see Fig. 1 of \citealp{2019ApJ...885..131R}). \citet{2016PASJ...68...60Y} measured the proper motion of G007.47+00.06 with VERA (VLBI Exploration of Radio Astrometry), and estimated a source distance of 20 $\pm$ 2 kpc based on the LSR velocity and proper motion in the direction of Galactic longitude (i.e., 2D kinematic distance). \citet{2017Sci...358..227S} confirmed the validity of the 2D kinematic distance via a trigonometric parallax distance measurement of the same source, which was derived to be 20.4$^{+2.8}_{-2.2}$ kpc with VLBA (Very Long Baseline Array). The reliability of 2D/3D kinematic distance was evaluated and confirmed for sources well past the Galactic center by \citet{2022arXiv220506903R}. Their work showed that 3D kinematic distances are more accurate than can be achieved with parallax measurements for distant sources. There is only one astrometric result (i.e., G007.47+00.06) in the outer portion of the Scutum-Centaurus arm while there are $\sim$40 astrometric results in the inner portion of the Scutum-Centaurus arm. To better understand the nature of the outer portion of the Scutum-Centaurus arm, it is necessary to increase the number of astrometric results for the outer portion of the arm.

The Extreme Outer Galaxy (EOG) is defined as the region outside the Outer spiral arm or at a Galactocentric distance $R \geq 2R_{0}$ where $R_{0}$ is the Galactocentric distance to the Sun (\citealp{1994ApJ...422...92D}). \citet{2011ApJ...734L..24D} found a new molecular arm beyond the Outer arm in the first Galactic quadrant and the arm is currently called the Outer-Scutum-Centaurus (OSC) arm. Also, \citet{2015ApJ...798L..27S, 2017ApJS..230...17S} discovered 72 EOG clouds (CO J = 1-0) in the second Galactic quadrant and 168 EOG clouds ($^{12}$CO and $^{13}$CO J = 1-0) in the Galactic longitude range 34.75$^{\circ} \leq l \leq$ 45.25$^{\circ}$. All the CO clouds detected by \citet{2015ApJ...798L..27S, 2017ApJS..230...17S} are associated with the OSC arm on the basis of a position-velocity diagram of CO (i.e., Galactic longitude $l$ vs. LSR velocity $v$). Excited-state OH (4.8 GHz), CH$_{3}$OH (6.7 GHz) and H$_{2}$O (22 GHz) masers were discovered toward brighter and massive EOG clouds by \citet{2018ApJ...869..148S}, although there was a limited number of detections (i.e., seven).

Here note that some of the EOG sources are close to the celestial equator $|\delta_{\rm{J2000}}|<9^{\circ}$, where making VLBI imaging is difficult. \citet{2011PASJ...63..513K} conducted VLBI astrometry toward a low-declination source, the star-forming region G034.43+00.24 ($\delta_{J2000}\sim$1$^{\circ}$) with VERA. Their imaging analysis suffered from side lobes of the synthesized interferometry beam pattern along the declination direction (see Fig. 7 of \citealp{2011PASJ...63..513K}), and they could use only astrometric results in the direction of right ascension for measuring the parallax of the source. To suppress the effect of such side lobes, increase in the number of antennas and an array configuration with minimum baseline redundancy are important. This is attributed to the fact that the width in $v$ direction is equal to that in the $u$ direction times the sine of the declination $\delta$ in the ($u$, $v$) coverage (\citealp{2017isra.book.....T}). If we observe a source with a declination of 0 degree using the four telescopes of VERA, we can obtain only six $v$ values. On the other hand, we can obtain 21 $v$ values when observing the same source with KaVA (KVN and VERA Array), where KVN is the Korean VLBI Network.

\begin{table*}[htbp] 
\caption{Observational information. \hspace{10em}} 
\begin{center} 
\label{table:4} 
\small 
\begin{tabular}{lllrlccc} 
\hline 
\hline 
Source&R.A. &Decl. 	  &Obs. Date	&Participating&Flux &\multicolumn{2}{c}{S/N}\\
\cline{7-8}	   &(J2000)	     &(J2000)		   & in UT&antenna&density &G034&J1851\\
	       &hh:mm:ss &dd:mm:ss &20yy/mm/dd&&(Jy/beam)\\ 
\hline
G034.84$-$00.95	&18:58:17.673  	&+01:16:06.40&A. 19/09/05&123$-$$-$ 67$-$$-$&1.6 &7 &75\\
 	&    	&&B. 19/10/11&12$-$ 45 67$-$$-$&1.5	&12 & 108\\       
 &&&C. 19/11/04	&12$-$ 45 67$-$$-$&2.0	&20&213	\\
 	&    	&&D. 19/12/21	&12 3 45 67$-$$-$	&1.9&23&208\\       
 	&&&E. 20/01/29&12 3 45 67$-$$-$		&2.2	&34 &271	\\   	&  	&&F. 20/02/27	&12 3 45 67$-$$-$	&1.9&34 &264\\              
 	&  	&&G. 20/03/22	&12 3 45 67$-$$-$&2.0	&29 &289\\              
&  	&&H. 20/04/19&12 3 45 \sout{6}\footnotemark[$\dag$]7$-$$-$&1.5&10&132\\
&  	&&I. 20/05/26 &1$-$3 45 67$-$$-$&$-$ & $<$ 5 &164 \\
&  	&&J. 20/09/22&1$-$3 45 6$-$ 89&0.6&5 &28\\
&  	&&K. 20/10/04&$-$23 4$-$$-$$-$89&$-$&$<$ 5&113 \\
&  	&&L. 20/11/12&1 23 4 5 6 7 89&2.4&21 &200\\
&  	&&M. 20/12/07&$-$23 4 5 6 7$-$$-$&$-$ &\multicolumn{2}{c}{Obs. failure}\\
	&  	&&N. 21/01/27&1 23 4 \sout{5}\footnotemark[$\ddag$] \sout{6}\footnotemark[$\ddag$] \sout{7}\footnotemark[$\ddag$] \sout{8}\footnotemark[$\ddag$]$-$&$-$ &$<$ 5 &96\\
	&    	&&O. 21/03/23&1 23 4 \sout{5}\footnotemark[$\ddag$] \sout{6}\footnotemark[$\ddag$] \sout{7}\footnotemark[$\ddag$] \sout{8}\footnotemark[$\ddag$]\sout{9}\footnotemark[$\ddag$]&2.2 &6 & 74\\
\hline 
\multicolumn{4}{@{}l@{}}{\hbox to 0pt{\parbox{160mm}{\footnotesize
\par\noindent
\\
Column 1 : 22 GHz H$_{2}$O maser source; 
Columns 2-3: equatorial coordinates in (J2000); Column 4: observation date in UT; Column 5: participating antennas (1 = Mizusawa; 2 = Iriki; 3 = Ogasawara; 4 = Ishigaki; 5 = Yonsei; 6= Ulsan; 7 = Tamna; 8 = Tianma; 9 = Nanshan); Column 6: flux density of the maser source; Columns 7-8: Signal to noise ratio values of the maser and a phase reference (i.e., J1851+0035) image where the maser is phase referenced to the continuum source.  \\ 
\footnotemark[$\dag$] Fringe of a bright calibrator was not detected in KVN Ulsan data.\\
\footnotemark[$\ddag$] Only VERA data could be used because fast antenna switching was not employed in the other stations. VERA joined the observations using the dual-beam system, and those observations were conducted for surveying new targets (see the text for details).\\

}\hss}}
\end{tabular} 
\end{center} 
\end{table*} 

\begin{figure*}[tbhp] 
 \begin{center} 
     \includegraphics[scale=0.91]{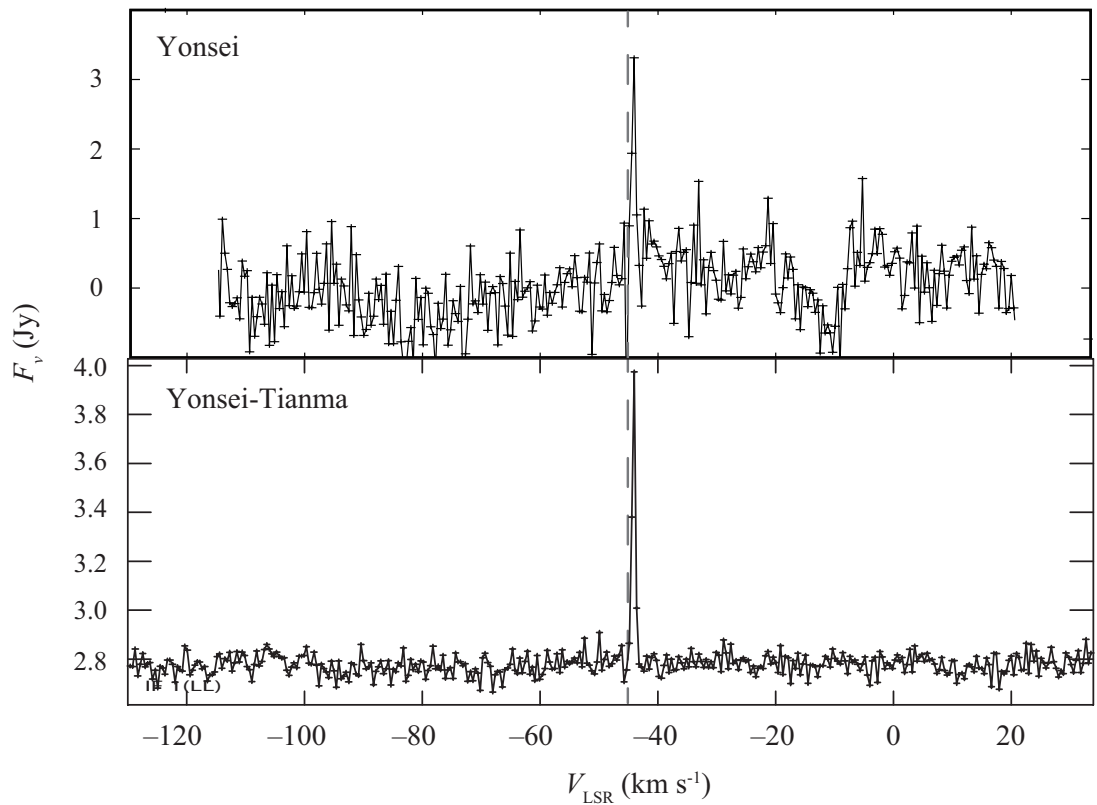} 
\end{center} 
\caption{ Total and cross-power spectra of G034.84$-$00.95 are shown with Yonsei {(\bf Top)} and Yonsei-Tianma baseline {(\bf Bottom)}, respectively. In the total power spectrum, data around both edges are not shown due to imperfect baseline subtraction. The observational data of epoch L is used (see Table \ref{table:4}). Vertical dashed lines indicate the systemic velocity of G034.84$-$00.95 which was determined by $^{12}$CO(J=1$-$0) observations ($-$45.1$\pm$2.2 km s$^{-1}$; \citealp{2017ApJS..230...17S}). }
\label{fig:12} 
\end{figure*}

 
In this paper, we report new astrometric results for H$_{2}$O masers associated with the star-forming region G034.84-00.95. G034.84$-$00.95 is associated with the OSC arm on the $l$-$v$ plot of CO, and the source emits H$_{2}$O (22.2 GHz) and OH (4.8 GHz) masers \citep{2018ApJ...869..148S}. The detection of OH masers suggests that the source is a more evolved star-forming region \citep{2022ApJS..260...51O}. \citet{2008AJ....136.2413R} classified the source as Young Stellar Object based on its infrared color [8.0] $-$ [24.0] $\geq$ 2.5 and magnitude [4.5] $>$ 7.8. However, no detailed research has been conducted for the source, besides survey results. 

We describe our observational setup and data reduction in sections 2 and 3, respectively. The astrometric results are shown in section 4. We discuss the validity of our astrometric results as well as the structure and kinematics of the OSC arm in section 5. In section 6, we summarize the paper.

\section{Observations}

Fifteen astrometric observations of 22 GHz H$_{2}$O masers associated with G034.84$-$00.95 were carried out between the 5th of September 2019 and the 23rd of March 2021 using KaVA which is part of the EAVN (East Asian VLBI Network)\footnote{see EAVN HP:\url{https://radio.kasi.re.kr/eavn/main_eavn.php}}. CVN (Chinese VLBI Network) telescopes Tianma 65m and Nanshan 26m have joined the KaVA astrometric observations since the 25th of May 2020. Observational information is shown in Table \ref{table:4}. We observed the maser source as well as phase references J1851+0035 $(\alpha, \delta)_{\rm J2000.0} =$ (\timeform{18h51m46s.7231}, $+$\timeform{00D35'32''.364}) and J1851+0035 $(\alpha, \delta)_{\rm J2000.0} =$ (\timeform{18h58m02s.3528}, $+$\timeform{03D13'16''.301}) for relative VLBI astrometry\footnote{The positions of background continuum sources were obtained from the Radio Fundamental Catalog of Astrogeo Center:\url{http://astrogeo.org/rfc/}}. Separation angles between the target and the phase references are 1.8$^{\circ}$ and 2.0$^{\circ}$, respectively. Note that J1858+0313 was observed in only eight epochs (i.e., epochs A$-$C and I$-$M in Table \ref{table:4}). While VERA used the dual beam system (\citealp{2000SPIE.4015..544K}) to simultaneously observe the target and a phase reference source, KVN employed fast antenna switching with a cycle of 1 (min) (20 seconds on target; 10 seconds antenna slewing; 20 seconds on a phase reference and a further 10 seconds antenna slewing). After the observational epoch ``I" in Table \ref{table:4}, KVN and CVN employed the fast antenna switching with a cycle of 80 (sec), except for observation epochs N and O in Table \ref{table:4}. We observed G034.84$-$00.95 as well as other sources (G039.18$-$01.43; G040.29+01.15; G040.96+02.48) in epochs N and O. Also, four half-hour ``geodetic blocks" spaced by about 2 hr were inserted in epochs between J and M for clock and atmospheric (tropospheric) delay calibration of the EAVN. For calibration of electric phase differences (i.e., manual phase calibration), a bright continuum source was observed for 5 (min) every $\sim$80 (min) in all observations.

Left-handed circular polarization data were recorded at 1024 Mbps with 2-bit quantization and the data were correlated with KJCC hardware correlator (\citealp{2015JKAS...48..125L}). Details of the back-end systems for KaVA and CVN are summarized in the status report of the EAVN\footnote{\url{https://radio.kasi.re.kr/status_report/files/status_report_EAVN_2022B.pdf}}. The maser data consisted of 16 MHz/1 IF and correlated with 512 channels, giving a frequency (velocity) spacing of 31.25 kHz (0.42 km s$^{-1}$) at a rest frequency of 22.235080 GHz (for H$_{2}$O maser). On the other hand, the continuum source data consisted of fifteen 16 MHz bands spanning 464 MHz where the fifteen IFs were distributed with an evenly spaced gap of 16 MHz, and each IF was correlated with 64 channels.

\begin{figure}[tbhp] 
 \begin{center} 
     \includegraphics[scale=0.9]{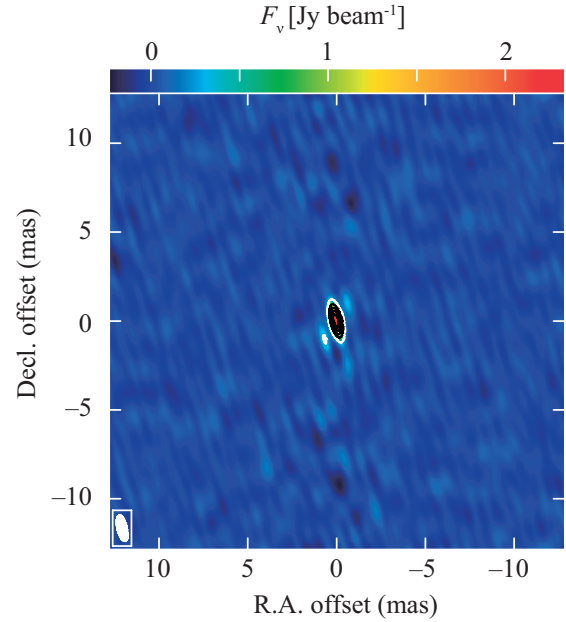} 
\end{center} 
\caption{ EAVN self-calibration image of G034.84$-$00.95 for observational epoch 'L' (see Table \ref{table:4}). Contour levels start at the 10$\sigma$ level and increase by factors of 10. The pattern of source's sidelobes, which is elongated from north-west to south-east directions, is due to the synthesized beam pattern.}
\label{fig:13} 
\end{figure}

\begin{figure}[tbhp] 
 \begin{center} 
     \includegraphics[scale=1.0]{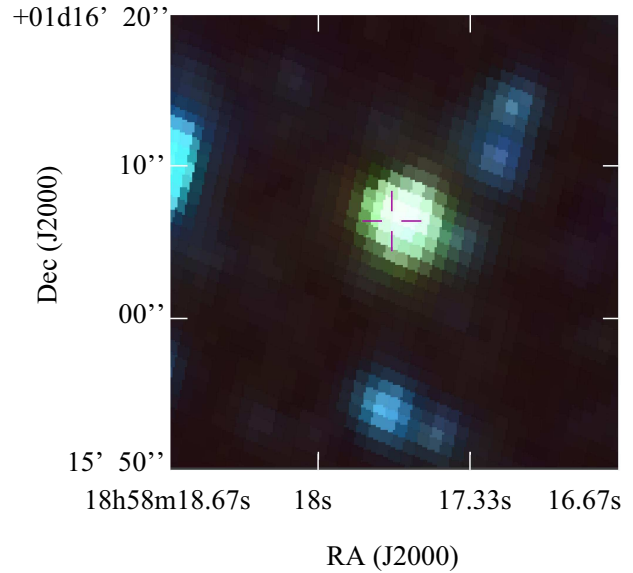} 
\end{center} 
 \caption{ $Spitzer$ RGB mid-infrared image (8.0, 4.5 and 3.6 $\mu$m) for the star-forming region G034.84$-$00.95 \citep{2003PASP..115..953B}. The position of the 22 GHz water maser associated with G034.84$-$00.95, is emphasized by the cross symbol.}
\label{fig:10} 
\end{figure}

\section{Data reduction}
Using the NRAO Astronomical Image Processing System ($AIPS$; \citealt{1996ASPC..101...37V}), data reduction was conducted with a standard procedure (e.g., see Fig. 11 of \citealp{2011PASJ...63..513K}). Since {\it a-priori} delay model used in the correlation processing was not accurate enough for high precision astrometry, the correlator model was replaced with a precise delay model consisting of an up-to-date geodynamical model (\citealp{1996ITN....21....1M}), the Earth orientation parameter EOP 08 C04 (IAU1980; C. Bizouard \& D. Gambis 2011\footnote{C. Bizouard \& D. Gambis 2011:\url{https://hpiers.obspm.fr/iers/eop/eopc04_08/C04.guide.pdf}}), station coordinates for KaVA antennas determined by Geodetic VLBI observations at K-band (the project's internal code v2005trf14; \citealp{2018JGSP...63...193}), ionospheric delays (GIM produced by CODE, the Center for Orbit Determination in Europe)\footnote{\url{ http://aiuws.unibe.ch/ionosphere/}}, and zenith wet excess path delays measured with GPS and the JMA (Japan Meteorological Agency meso-scale analysis data for numerical weather prediction\footnote{\url{ https://www.jma.go.jp/jma/jma-eng/jma-center/nwp/outline2019-nwp/index.htm}}). We applied GPS for VERA because the time resolution of GPS (5 min) was shorter than that of JMA (3 hr). On the other hand, we applied JMA-based corrections to KVN data since GPS data were unavailable during the observational period. Note that \citet{2015PASJ...67...65N} demonstrated that tropospheric zenith delay residual ($c\Delta \tau_{\rm{trop}}$) can be suppressed within $\sim$2 cm using either GPS or JMA.

The effect of the time variation of the parallactic angles of the telescope was corrected with the AIPS task $\texttt{CLCOR}$ with the adverbs $\texttt{OPCODE = 'PANG'}$ and $\texttt{CLCORPRM(1)=1}$, except for VERA  where this effect does not appear thanks to the field rotator equipped for the VERA dual-beam system. As explained previously, KVN and CVN observed the maser source and a phase reference in fast antenna switching, whereas VERA observed both the sources simultaneously using the dual-beam system. Thus, we made a flag file (FG table) with the AIPS task $\texttt{UVFLG}$, which was applied for the antenna slew time of the KVN and CVN stations.

\begin{figure*}[tbhp] 
 \begin{center} 
     \includegraphics[scale=1.0]{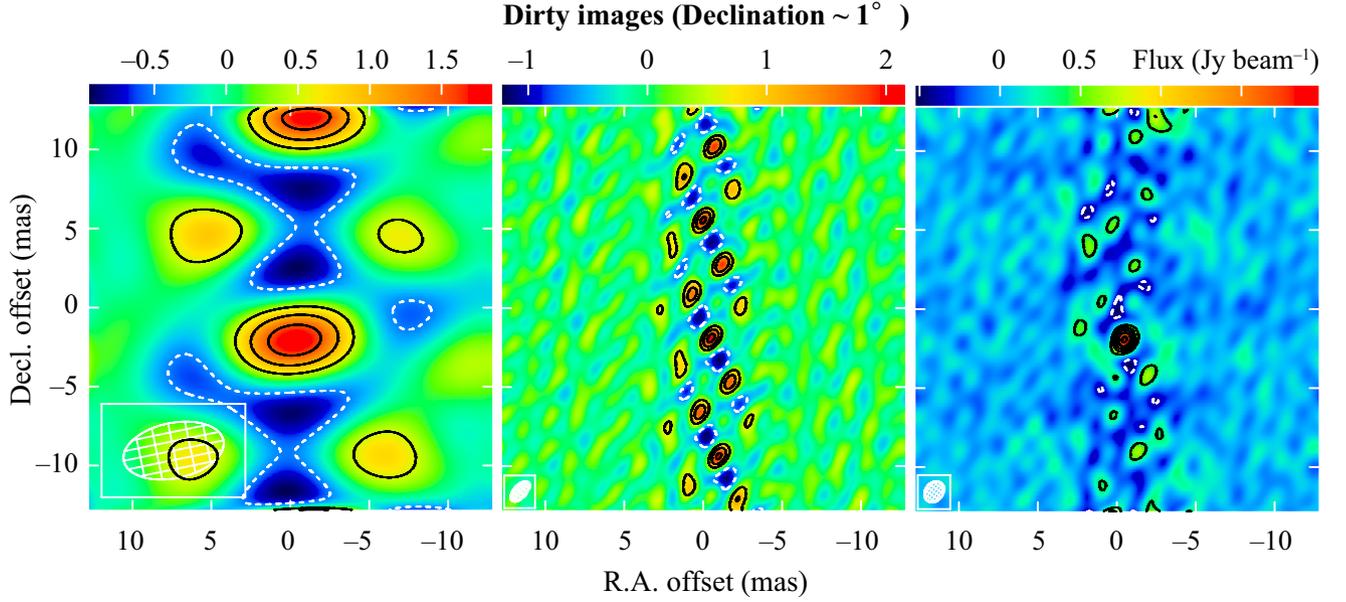} 
\end{center} 
\caption{ Dirty images of G034.84$-$00.95 obtained with KVN {\bf(Left)}, VERA {\bf(Middle)}, and KaVA {\bf(Right)}, respectively. The observation was conducted on the 22nd of March 2020 (observation epoch G in Table \ref{table:4}). Solid lines show contour levels of the images with a step of 3$\sigma$ noise level. The dashed lines are the same as the solid ones, but with a step of $-$3$\sigma$.}
\label{fig:7} 
\end{figure*}

\begin{figure*}[tbhp] 
 \begin{center} 
     \includegraphics[scale=1.28]{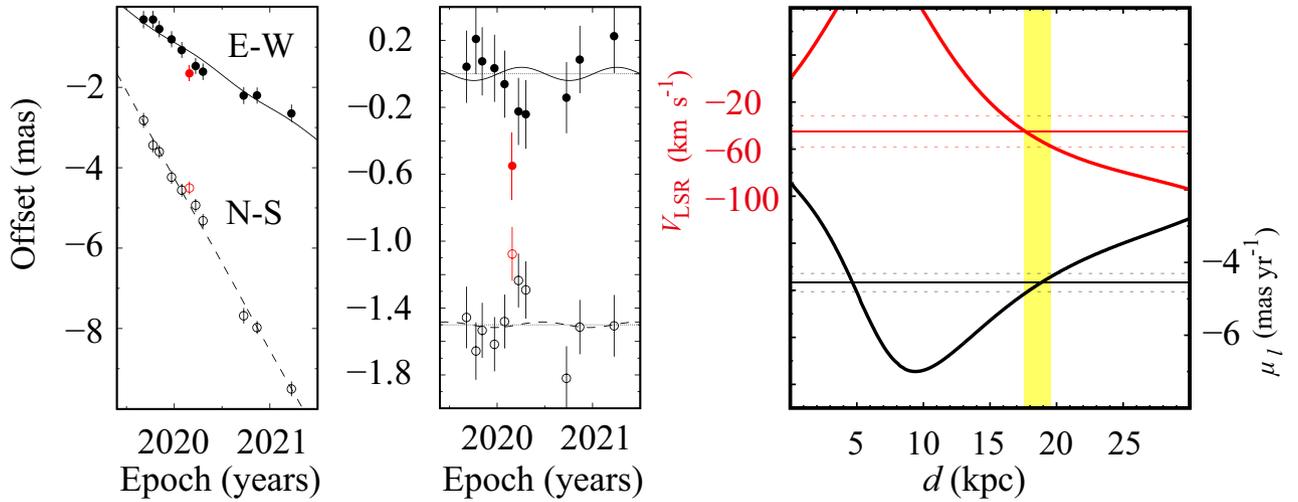} 
\end{center} 
\caption{ {\bf(Left)} Results of parallax and proper motion fittings for G034.84-00.95. The position offsets of the masers are plotted with respect to the background QSO J1851+0035 in the east ($\alpha$cos$\delta$) and north ($\delta$) directions as a function of time. For clarity, the northerly data is plotted with an offset from the easterly data. The best-fit models in the easterly and northerly directions are shown as continuous and dashed curves, respectively. The data points denoted by red circles are removed from the fitting using the ``conservative formulation'' of \citealp{2006book...58..Sivia} (see Appendix \ref{appendix:2} for details).  {\bf(Middle)} Same as the left, but with the fitted proper motion removed. 
 {\bf(Right)} Kinematic distance models through LSR velocity $V_{{\rm LSR}}$ (red curve; Y1 axis) and the proper motion in the direction of Galactic longitude $\mu_{l}$ (black curve; Y2 axis). The horizontal axis is the heliocentric distance. Here, we assume the Galactic constants to be ($R_{0}$ = 8.15 kpc, $\Theta_{0}$ = 236 km s$^{-1}$) and a universal rotation curve (A5 model of \citealp{2019ApJ...885..131R}). Horizontal red and black lines present the observed best values of $V_{{\rm LSR}}$ and $\mu_{l}$, respectively. Dotted red lines represent the error range where $\pm$13 km s$^{-1}$ is added in quadrature to a formal error of $V_{{\rm LSR}}$, while dotted black lines are the same as the red ones but for $\mu_{l}$ (see the text for details). The yellow shaded region indicates the range for the weighted average of the two kinematic distances.}
\label{fig:1} 
\end{figure*}  

\begin{table*}[htbp] 
\caption{Parallax, proper motion, and kinematic distance results for G034.84$-$00.95. \hspace{10em}} 
\begin{center} 
\label{table:2} 
\small 
\begin{tabular}{lcccccccc} 
\hline 
\hline 
 Target                 &Parallax ($\pi$)\footnotemark[*] &$\mu_{\alpha} \rm{cos}\delta$       &$\mu_{\delta}$&$V_{\rm{LSR}}$    &\multicolumn{3}{c}{Kinematic distance}      &Ref. \\ \cline{6-8}
   
                        &   & & & &$d_{V_{\rm{LSR}}}$ &$d_{\mu_{l}}$ &$d_{\rm{2D}}$ \\
        &(mas)&(mas yr$^{-1}$)    &(mas yr$^{-1}$)&(km s$^{-1}$)& (kpc)&(kpc)&(kpc)                                          \\
  \hline
G034.84$-$00.95                                   &0.040$^{+0.061}_{-0.028}$                                   &$-$1.61$\pm$0.18        &$-$4.29$\pm$0.16              &$-$45.1$\pm$2.2&17.7$^{+2.0}_{-1.6}$&18.9$\pm$1.1&18.6$\pm$1.0&a                     \\

\hline 
\multicolumn{4}{@{}l@{}}{\hbox to 0pt{\parbox{160mm}{\normalsize
\par\noindent
\\
Column 1: 22 GHz H$_{2}$O maser source; Columns 2: parallax; Columns 3-4; proper motion components in east and north directions, respectively; Column 5: LSR velocity; Columns 6-7: kinematic distances through LSR velocity and proper motion in the direction of Galactic longitude, respectively (see the text for details); Column 8: the weighted average of both Columns 6 and 7; Column 9: reference for the LSR velocity estimated from a molecular line observation. \\
$\bf{Reference}$: (a) \citet{2017ApJS..230...17S}. \\
\footnotemark[*] Since the fractional parallax error is significantly greater than 20$\%$, distance estimation by simply inverting the parallax results in a significant bias (see \citealp{2015PASP..127..994B}).   \\
}\hss}}
\end{tabular} 
\end{center} 
\end{table*} 

After standard instrumental delay calibration, phase-referencing procedures, and source image synthesis, the position of the maser source relative to a phase reference source was determined by an elliptical Gaussian fitting with the AIPS task ``JMFIT" in each observation epoch. Regarding the phase-referencing procedure, we used solution derived for a nearby continuum source to phase reference the maser source. To guarantee the reliability of the data reduction (i.e., phase referencing), we show corrected phases in Fig. \ref{fig:11} of Appendix \ref{appendix:1}. 

Maser positions were recorded as a function of time and modeled by (1) an annual parallax, (2) linear proper motion components in east--west ($\alpha$cos$\delta$) and north--south ($\delta$) directions, and (3) reference positions in $\alpha$cos$\delta$ ($\alpha_{0}$) and $\delta$ ($\delta_{0}$) (e.g., see equations 1-2 in \citealp{2012PASJ...64....7K}). The model parameters were determined with a Bayesian approach (see Appendix \ref{appendix:2} for details) where a systematic error was added in quadrature to a formal (thermal) error so that the reduced chi-square value became nearly unity. This is because systematic error caused by the tropospheric zenith delay residual is dominant in 22 GHz VLBI astrometry for distant sources (e.g., \citealp{2020PASJ...72...52N}).

\section{Results}
\label{Results}

\subsection{EAVN self-calibration results}
\label{self-calibration}
Figure \ref{fig:12} shows total and cross-power spectra of the 22 GHz water maser in G034.84$-$00.95. We confirm the single peak emission whose velocity is consistent with the peak velocity of $^{12}$CO (J=1$-$0) ($-$45.1$\pm$2.2 km s$^{-1}$; \citealp{2017ApJS..230...17S}). Figure \ref{fig:13} displays an EAVN self-calibration image of G034.84$-$00.95 for the observational epoch 'L' (see the corresponding date in Table \ref{table:4}). We could make an image only for the peak emission. We confirmed that the signal to noise (S/N) ratio of the EAVN image was 1.3 times increased (46$\rightarrow$60) compared to that of the KaVA image even though Tianma 65m and Nanshan 25m telescopes participated for only $\sim$3 hours (roughly one third of the total observing time) in observational epoch L. We also confirmed that the maser spot was not resolved with the longest baseline length of 5200 km in Fig. \ref{fig:13}. The maser spot is compact and suitable for parallax and proper-motion fitting. \\

\subsection{KaVA astrometry results}
\label{astrometry}

We report only KaVA astrometric results in this paper since zenith wet delay calibration of EAVN (i.e., KaVA, Tianma 65m, and Nanshan 26m) based on the Geodetic blocks remains under evaluation. 
We successfully produced maser images with phase referencing in 11 out of 15 observations (see Table \ref{table:4}). The position of the maser source is superimposed on a $Spitzer$ RGB mid-infrared image (Fig. \ref{fig:10}) where the angular resolution of the $Spitzer$ image is 3 (arcsec) and the position of the maser source is consistent with that of the Spitzer/GLIMPSE source G034.8427-00.9465 within 0.4 (arcsec) \citep{2009yCat.2293....0S}. Failures of the rest of the observations are due to insufficient signal to noise ratio (S/N $<$ 5) of the phase-referenced (i.e., maser) image and incorrect frequency setup. We found correlation between S/N ratio values of images of the maser and a phase reference J1851+0035 with a correlation coefficient of 0.96 (see Table \ref{table:4}). 

Figure \ref{fig:7} demonstrates that the side lobe along the north-south direction can be well suppressed in the KaVA dirty image. We confirmed that correlated flux density of the maser source was $\sim$2 Jy beam$^{-1}$ at a line-of-sight velocity of $V_{\rm{H}_{2}O}$ $\sim$ $-$44 km s$^{-1}$ during the observations. The velocity of G034.84$-$00.95 is consistent with the LSR velocity of the source obtained by $^{12}$CO(J=1$-$0) observations ($-$45.1$\pm$2.2 km s$^{-1}$; \citealp{2017ApJS..230...17S}). 

We measured the proper motion components of the maser spot to be ($\mu_{\alpha} \rm{cos}\delta$, $\mu_{\delta}$) = ($-$1.61$\pm$0.18,$-$4.29$\pm$0.16) mas yr$^{-1}$ in equatorial coordinates, however we could not obtain a reliable result of the parallax fitting (Table \ref{table:2} and Fig. \ref{fig:1}). Since H$_{2}$O masers are typically generated in outflows of tens of km s$^{-1}$ and transferring the motion of the maser to that of the central star causes an uncertainty, we adopted a measurement uncertainty of $\pm$10 km s$^{-1}$ in quadrature to each motion component in Table \ref{table:2} by considering the fact that we detected only one maser spot throughout the observations. The systematic error of 10 km s$^{-1}$ is converted to 0.11 mas yr$^{-1}$ at a source distance of 18.9 kpc which is determined by the measured proper motion as explained below.

\begin{figure*}[tbhp] 
 \begin{center} 
     \includegraphics[scale=0.8]{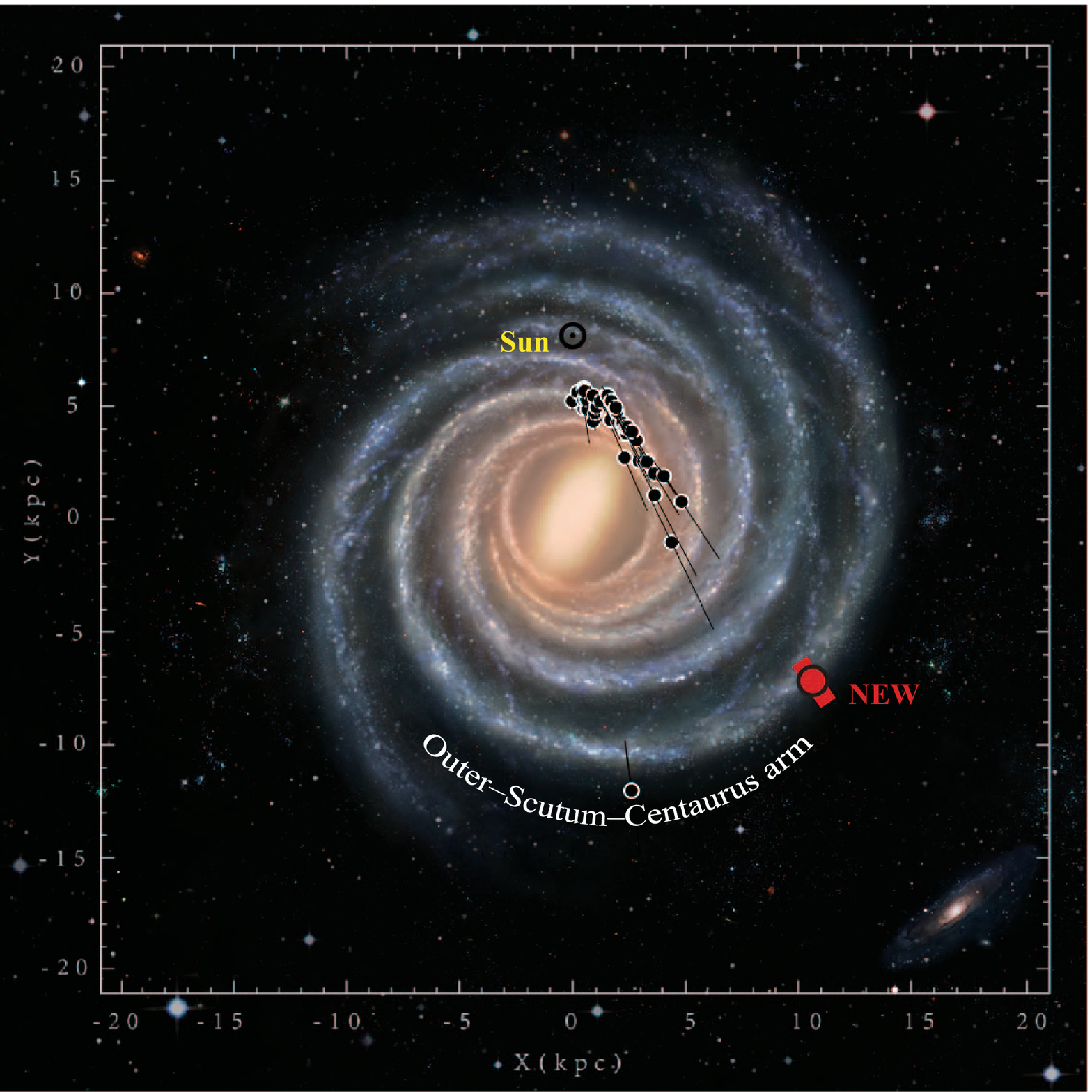} 
\end{center} 
 \caption{ The most scientifically accurate visualization of the Milky Way (at the moment) on which previous VLBI astrometric results (black circles with white borders; \citealp{2019ApJ...885..131R}; \citealp{2020PASJ...72...50V}) for the Scutum-Centaurus arm are superimposed. The new astrometric result of G034.84$-$00.95 ($d$ = 18.6$\pm$1.0 kpc) is emphasized by the red circle with black border. {\bf Image Credit\protect: }\copyright Xingwu Zheng $\&$ Mark Reid, BeSSeL/NJU/CfA. The background image can be downloaded from the URL\protect\footnotemark.}
\label{fig:2} 
\end{figure*}

We estimated kinematic distances of the source through LSR velocity $V_{\rm{LSR}}$ and the proper motion in the direction of Galactic longitude $\mu_{l}$ by referring to \citet{2022PASJ...74..209S} (see also Table \ref{table:2}). For estimating the kinematic distances, we assumed the solar motion to be ($U_{\odot}$, $V_{\odot}$, $W_{\odot}$) = (10.6, 10.7. 7.6) km s$^{-1}$, the Galactic constants to be ($R_{0}$, $\Theta_{0}$) = (8.15 kpc, 236 km s$^{-1}$), and a universal rotation curve (i.e., A5 model of \citealp{2019ApJ...885..131R}).
To allow the effect of noncircular motion caused by the spiral arm, we added 13 km s$^{-1}$ in quadrature to the formal error of $V_{\rm{LSR}}$. The uncertainty of 13 km s$^{-1}$ corresponds to the mean noncircular motion of the outer Perseus arm \citep{2019ApJ...876...30S}. The error range of $\mu_{l}$ was estimated in the same procedure. The resultant kinematic distances, $d_{V_{\rm{LSR}}}$ and $d_{\mu_{l}}$, are 17.7$^{+2.0}_{-1.6}$ kpc and 18.9$\pm$1.1 kpc, respectively. The kinematic distance $d_{\mu_{l}}$ is more accurate than $d_{V_{\rm{LSR}}}$ in the case of G034.84$-$00.95, since the accuracy distributions of $d_{\mu_{l}}$ and $d_{V_{\rm{LSR}}}$ are different on the Galactic coordinates (see \citealp{2011PASJ...63..813S}). The two dimensional kinematic distance, $d_{{\rm 2D}}$, as the variance weighted average of both $d_{V_{\rm{LSR}}}$ and $d_{\mu_{l}}$, is 18.6$\pm$1.0 kpc. We further examined the validity of our distance estimation in Appendix \ref{appendix:3} where 2D kinematic distances are determined by shifting LSR velocity and the measured proper motion individually by 40 km s$^{-1}$. We confirmed that all the 2D kinematic distances are consistent within errors. Note that the validity of our distance estimate should be checked by measuring the trigonometric parallax of G034.84$-$00.95 in the future. The 2D kinematic distance (i.e., $d_{2D}$ = 18.6$\pm$1.0 kpc) places the source in the Outer-Scutum-Centaurus arm (Fig. \ref{fig:2}).

Note that the naming of ``2D distance'' comes from the fact that we didn't consider the proper motion perpendicular to the disk (i.e., $\mu_{b}$) in the distance estimation of G034.84$-$00.95, because the vertical motion is a weak constraint compared to other motion components $V_{\rm{LSR}}$ and ${\mu_{l}}$ in our case. To verify the above explanation, we used ``A Parallax-based Distance Calculator\footnote{\url{http://bessel.vlbi-astrometry.org/node/378}}'' (\citealp{2016ApJ...823...77R,2019ApJ...885..131R}) where ``3D motion'' is considered to estimate the distance of a source. We confirmed that the distance constraint by the vertical motion $\mu_{b}$ has a large tail in the probability density function of the distance. The resultant distance of the calculator is 18.5$\pm$1.0 kpc with a probability of 0.89, which is consistent with our independent estimate (18.6$\pm$+/-1.0 kpc) within errors.

\section{Discussion}
\subsection{Is G034.84-00.95 really a distant object ?}

\begin{figure*}[tbhp] 
 \begin{center} 
     \includegraphics[scale=0.95]{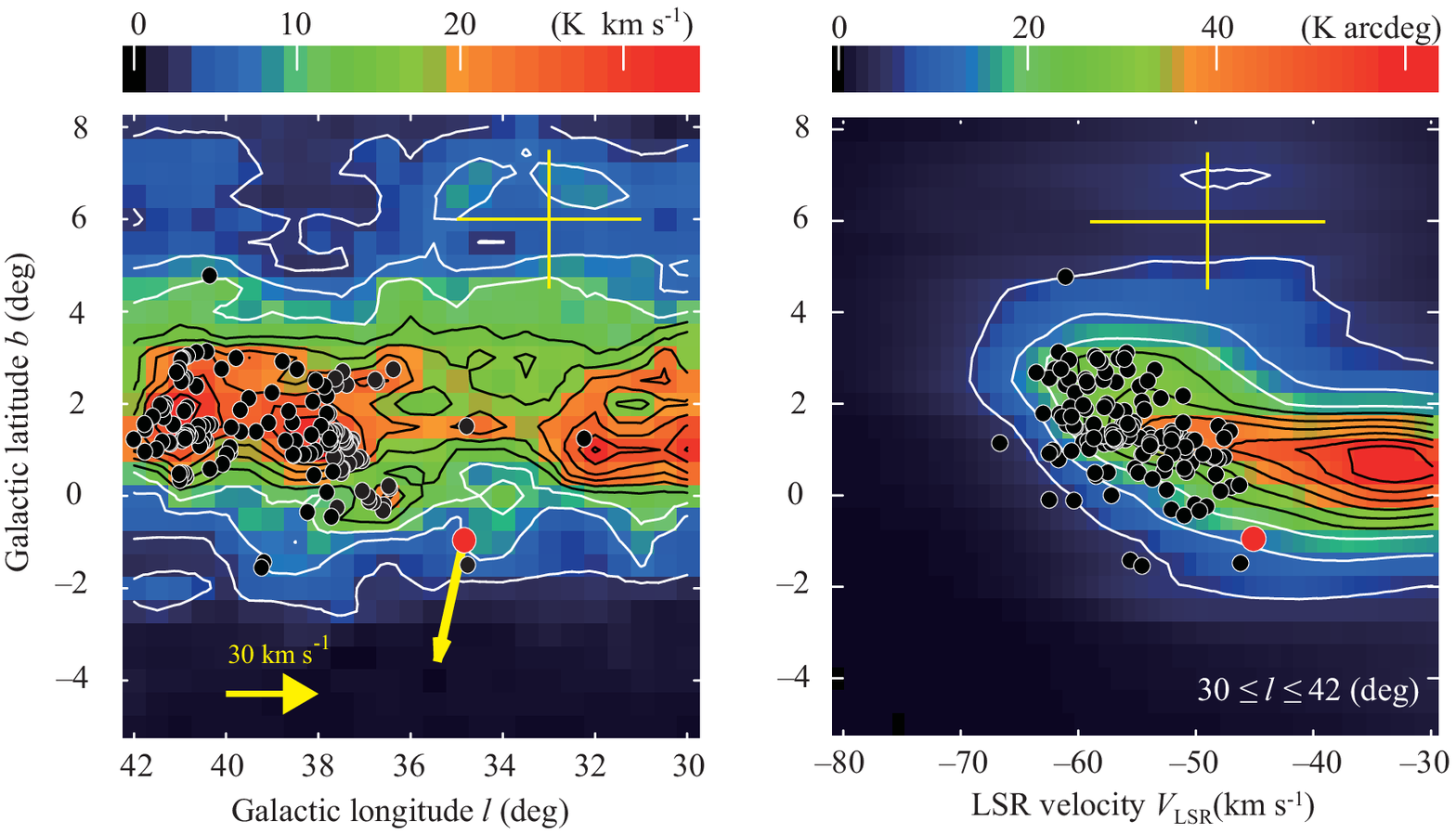} 
\end{center} 
\caption{ {\bf(Left)} Velocity-integrated H {\tiny I} emission in Galactic coordinates, obtained by integrating LAB H {\tiny I} survey data \citep{2005A&A...440..775K}.
The integration range of LSR velocity is consistent with the Outer-Scutum-Centaurus arm (see Fig. 2 of \citealp{2017ApJS..230...17S} for the range). The H {\tiny I} contour starts at 3.4 (K km s$^{-1}$) and increases by a factor of $n$ where $n$ = 2,3,,,8,9. Circles show CO clouds associated with the Outer-Scutum-Centaurus arm, taken from \citet{2011ApJ...734L..24D} and \citet{2017ApJS..230...17S}. Red circle represents G034.84$-$00.95, while the yellow arrow displays the noncircular motion of the source. A yellow cross shows the scale of the H {\tiny I} supershell GS033+06$-$49 located at a heliocentric distance of $\sim$22 kpc (\citealp{1979ApJ...229..533H}). {\bf(Right)} Same as the Left, but for LSR velocity vs. Galactic latitude of the H {\tiny I} emission, integrated over the Galactic longitude range 30 $\leqq l \leqq$ 42 (deg). The H {\tiny I} contour starts at 6.3 (K arcdeg) and increases by a factor of $n$ where $n$ = 2,3,,,8,9. }
\label{fig:4} 
\end{figure*}

\begin{figure}[tbhp] 
 \begin{center} 
     \includegraphics[scale=1.0]{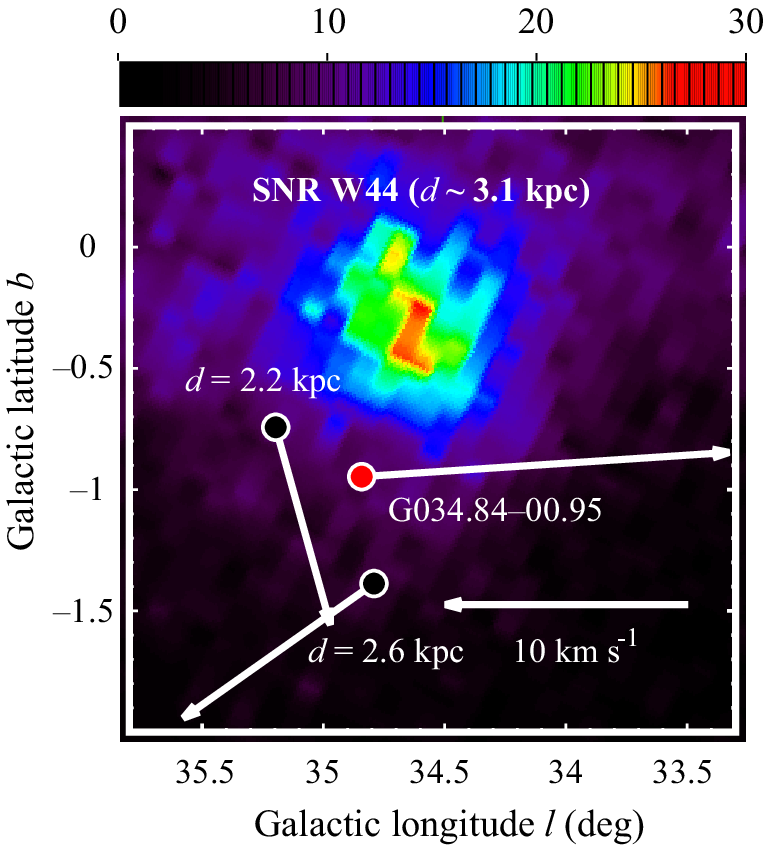} 
\end{center} 
\caption{ Fermi-LAT (1$-$3 Gev) intensity map \citep{2009ApJ...697.1071A}\protect\footnotemark[9] of the supernova remnant W44 on which VLBI astrometric results are superimposed. The intensity is shown by colors in units of (photon cm$^{-1}$ s$^{-1}$ sr$^{-1}$). The vectors show noncircular motion of the maser sources derived from the VLBI astrometric results, referring to the Case A of Table \ref{table:5} where the heliocentric distance to G034.84$-$00.95 (red circle) is assumed to be 2.5$\pm$0.1 kpc. }
\label{fig:3} 
\end{figure}

Although most CO clouds associated with the Outer-Scutum-Centaurus arm show positive Galactic latitude values at the Galactic longitude range 30$^{\circ} \leq l \leq$ 42$^{\circ}$, G034.84$-$00.95 has a negative Galactic latitude (see Fig. \ref{fig:4}). This indicates that the source may be a nearby object, which is inconsistent with our distance estimate of the source ($d$ = 18.6$\pm$1.0 kpc). Indeed, nearby star-forming regions G034.79$-$01.38 ($d$ = 2.62$^{+0.15}_{-0.13}$ kpc; \citealp{2019ApJ...885..131R}) and G035.19$-$00.74 ($d$ = 2.19$^{+0.24}_{-0.20}$ kpc; \citealp{2019ApJ...885..131R}) as well as the supernova remnant W44 ($d\sim$3.1 kpc; \citealp{2014A&A...565A..74C}) are observed toward similar directions of G034.84$-$00.95 (see Fig. \ref{fig:3}). However, we emphasize that the parallax signature corresponding to 2.5 kpc (i.e., 0.4 mas), should be detected with KaVA if the source is located at $d$ = 2.5 kpc. The small parallax result with relatively large uncertainty (0.040$^{+0.061}_{-0.028}$ mas) supports a larger distance. To confirm the validity of our parallax result quantitatively, we superimposed a larger parallax amplitude of 0.4 mas on our parallax result (see Fig. \ref{fig:9} in Appendix \ref{appendix:1}) and compared reduced chi-square values between both parallax results. The reduced chi-square value is increased from 1.1 to 2.9 if we apply the larger 0.4 mas parallax amplitude. Thus, the observational results can be better explained with the smaller parallax amplitude; i.e. the far distance. 

\footnotetext[8]{\url{https://astronomy.nju.edu.cn/xtzl/EN/index.html}}

\begin{table*}[htbp] 
\caption{Noncircular motion and spectral type of G034.84$-$00.95 at different distances. \hspace{10em}} 
\begin{center} 
\label{table:5} 
\small 
\begin{tabular}{lcccclclll} 
\hline 
\hline
Source& $U_{s}$         & $V_{s}$ 	   & $W_{s}$	  & $d$  & $R$  & $z$ &log$L$/$L_{\odot}$ &Spectral\\
      &(km s$^{-1}$)  &(km s$^{-1}$)      &(km s$^{-1}$) &(kpc) &(kpc) &(pc) &&type \\ 
\hline
 \multicolumn{3}{l}{{\bf Case A: Distance = 2.5$\pm$0.1 kpc}\footnotemark[$*$]}		\\

 G034.84$-$00.95	&$-$48$\pm$4  	&$-$76$\pm$4 &1$\pm$2 &2.5$\pm$0.1 &6.3$\pm$0.1      &$-$42$\pm$2&2.99&Later than B3	 \\
 G034.79$-$01.38	&$-$7$\pm$7  	&4$\pm$7 &$-$6$\pm$8 &2.6$\pm$0.1 &6.2$\pm$0.1      &$-$64$\pm$3&3.71&Earlier than B2	 \\
 G035.19$-$00.74	&$-$4$\pm$7  	&$-$8$\pm$7 &$-$8$\pm$5 &2.2$\pm$0.2 &6.5$^{+0.1}_{-0.2}$ &$-$29$\pm$3&4.28&Earlier than B0.5\\	 \\

  \multicolumn{3}{l}{{\bf Case B: Distance = 18.6$\pm$1.0 kpc}\footnotemark[$\dag$]}		\\

 G034.84$-$00.95	&$-$6$\pm$10  	&$-$7$\pm$24 &$-$38$\pm$16 &18.6$\pm$1.0 &12.8$^{+1.0}_{-0.9}$      &$-$309$\pm$17&4.73&O8.5	 \\

 \hline 
\multicolumn{4}{@{}l@{}}{\hbox to 0pt{\parbox{165mm}{\footnotesize
\par\noindent
\\
Column 1 : source name; 
Columns 2-4: noncircular motion components toward the Galactic center, in the direction of the Galactic rotation, and toward the north Galactic pole, respectively; Column 5: heliocentric distance; Column 6: Galactocentric distance; Column 7: Galactic height; Column 8: Bolometric luminosity estimated from infrared flux densities in four bands (see \citealp{2009A&A...507..369W}); Column 9: Spectral type of Zero Age Main Sequence (see \citealp{1973AJ.....78..929P}). \\
\footnotemark[$*$] Variance weighted average of the distances of G034.79$-$01.38 and G035.19$-$00.74. \\
\footnotemark[$\dag$] Our result (see the text for details). \\

}\hss}}
\end{tabular} 
\end{center} 
\end{table*} 

To further evaluate the possibility that the source is a nearby object, we estimate peculiar (i.e., noncircular) motion and the spectral type of G034.84$-$00.95 at different distances (2.5$\pm$0.1 kpc and 18.6$\pm$1.0 kpc) in Table \ref{table:5}. We also derive the physical quantities of G034.79$-$01.38 and G035.19$-$00.74 for comparison. Note that the former distance of G034.84$-$00.95 is the variance weighted average of the distances of G034.79$-$01.38 and G035.19$-$00.74. The calculation of noncircular motion values is based on the appendices of \citet{2015PASJ...67...69S,2020PASJ...72...53S}. For the derivation of bolometric luminosity, we use flux densities in IRAS four bands (12 $\mu$m, 25 $\mu$m, 60 $\mu$m, and 100 $\mu$m) and the distance for each object (see the formulation by \citealp{2009A&A...507..369W}). We independently estimated the bolometric luminosity of G034.84$-$00.95 with Spectral Energy Distribution fits in Appendix \ref{appendix:4}. We confirmed that both the methods can provide consistent results to each other. 

If we assume a distance of 2.5$\pm$0.1 kpc for G034.84$-$00.95, a significantly large noncircular motion is derived along the disk (i.e., $U_{s}$ = $-$48$\pm$4 km s$^{-1}$ and $V_{s}$ = $-$76$\pm$4 km s$^{-1}$). However, such noncircular motion is not seen for G034.79$-$01.38 and G035.19$-$00.74 (see Table \ref{table:5}). Backwinding the noncircular motion of G034.84$-$00.95 does not intersect the SNR W44 in Fig. \ref{fig:3}. The bolometric luminosity of the source is less luminous (2.99 in log$L$/$L_{\odot}$) than the other sources. Thus, the near distance (2.5$\pm$0.1 kpc) is unlikely for G034.84$-$00.95.

Our distance estimate for G034.84$-$00.95 (18.6$\pm$1.0 kpc) is also supported by the fact that a Galactic H {\scriptsize II} region associated with the OSC arm shows negative Galactic latitude at ($l$, $b$, $V_{\rm{LSR}}$) = (39$\fdg$183, $-$1$\fdg$422, $-$54.9 km s$^{-1}$) \citep{2015ApJS..221...26A}. If our distance estimate is correct, G034.84$-$00.95 is regarded as a high-mass star-forming region with a bolometric luminosity of 4.73 (in log$L$/$L_{\odot}$) and a spectral type of O8.5. Also, the source is located at a Galactic height $z$ of $-$309$\pm$17 pc (see Table \ref{table:5}). The source is consistent with the Galactic height range ($-$500 pc $< z < $ 1500 pc) of Extreme Outer Galaxy CO clouds \citep{2017ApJS..230...17S}. The source would show large noncircular motion perpendicular to the disk ($W_{s}$ = $-$38$\pm$16 km s$^{-1}$) at a distance of 18.6$\pm$1.0 kpc. We will discuss a possible origin of the vertical motion in the next subsection.

\begin{figure}[tbhp] 
 \begin{center} 
     \includegraphics[scale=0.95]{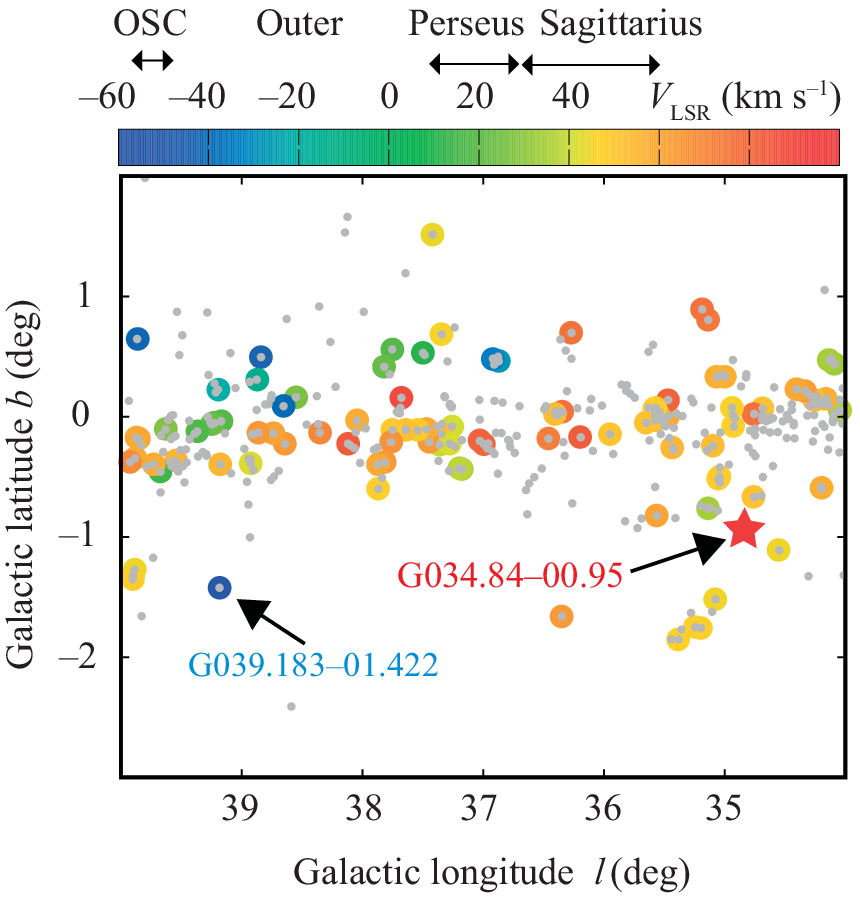} 
\end{center} 
\caption{ Spatial distribution of Galactic H {\tiny II} regions in Galactic coordinates (gray circles; \citealp{2014ApJS..212....1A, 2015ApJS..221...26A}). G034.84$-$00.95 indicated by the red star, is superimposed on the plot. Color displays LSR velocity for a part of the H {\tiny II} regions. Horizontal arrows on the color bar show expected velocity ranges for individual spiral arms based on \citet{2017ApJS..230...17S} and \citet{2019ApJ...885..131R}. Note that OSC is the Outer-Scutum-Centaurus arm. The expected velocity range is not shown for the Outer arm, since no VLBI astrometric results for the arm are available in the Galactic longitude range $l$ = [34$^{\circ}$:40$^{\circ}$]. Black arrows emphasize G034.84$-$00.95 and the H {\tiny II} region G039.183$-$01.422 which are associated with the OSC arm.} 
\label{fig:6} 
\end{figure}  

\footnotetext[9]{Downloaded using the software \texttt{Aladin Sky Atlas} \citep{2000A&AS..143...33B,2014ASPC..485..277B}.}

\subsection{Vertical motion perpendicular to the disk}

Fast vertical motion ($V_{z}\geq$ 20 km s$^{-1}$) has been discussed in the context of the expansion of the NGC281 superbubble (\citealp{2008PASJ...60..975S}), whereas \citet{2022PASJ...74..209S} discussed it in another context as originating in cloud collisions with the Galactic disk. The H {\scriptsize I} supershell GS033+06$-$49 exhibits a Galactic longitude and LSR velocity similar to G034.84$-$00.95 (see Fig. \ref{fig:4}), and a kinematic distance of $\sim$ 22 kpc was estimated for the H {\scriptsize I} supershell (\citealp{1979ApJ...229..533H}). If we assume the same rotation curve for G034.84$-$00.95 and the H {\scriptsize I} supershell, then the difference of individual distances between the two objects should become smaller. The existence of the H {\scriptsize I} supershell indicates that gas circulation occurs between the outer Galactic disk and halo at a Galactocentric distance $>$12 kpc in the first Galactic quadrant. Note that a physical relationship between G034.84$-$00.95 and the H {\scriptsize I} supershell is unclear, because G034.84$-$00.95 is deviated from the H {\scriptsize I} supershell by $\sim$7$^{\circ}$ in the direction of Galactic latitude.

\begin{figure}[tbhp] 
 \begin{center} 
     \includegraphics[scale=0.95]{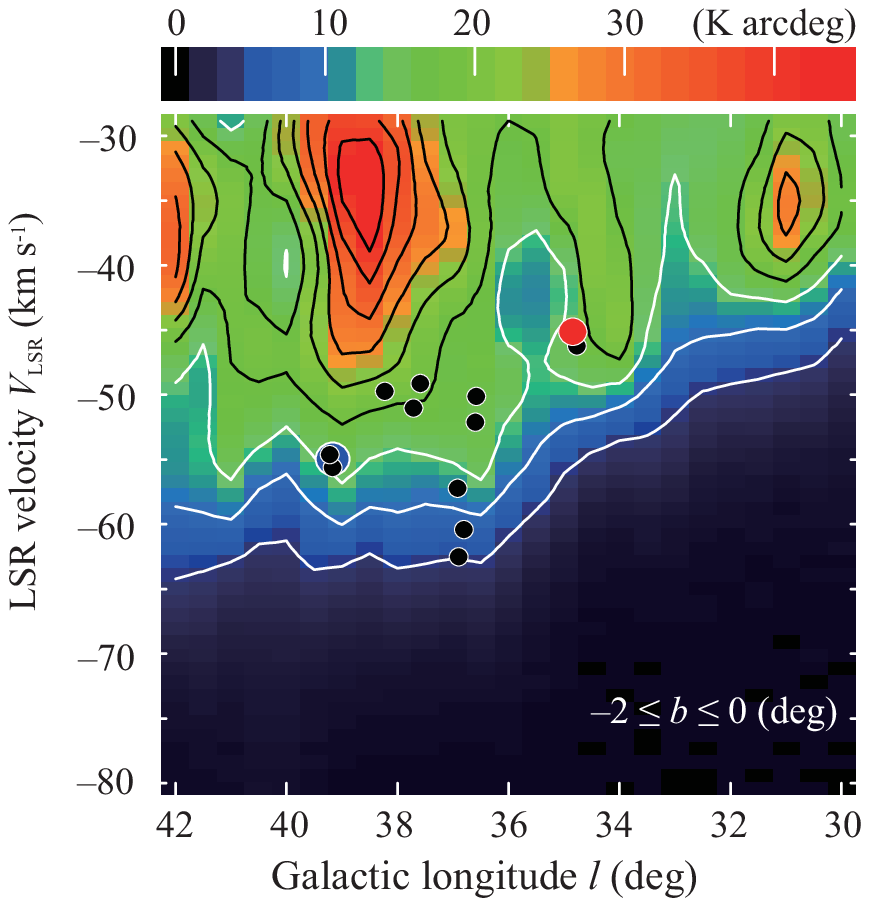} 
\end{center} 
\caption{ Longitude-velocity diagram of H {\tiny I} emission, obtained by integrating LAB H {\tiny I} survey data from \citet{2005A&A...440..775K} over the latitude range $-$2 $\leqq b \leqq$ 0 (deg). The H {\tiny I} contour starts at 4.4 (K arcdeg) and increases by a factor of $n$ where $n$ = 2,3,,,8,9. Black circles show CO clouds associated with the OSC arm, taken from \citet{2011ApJ...734L..24D} and \citet{2017ApJS..230...17S}. Red circle represents G034.84$-$00.95, while the blue one shows the H {\tiny II} region G039.183$-$01.422.}
\label{fig:5} 
\end{figure}

We check the distribution of H {\scriptsize II} regions around G034.84$-$00.95, since the NGC281 superbubble is surrounded by an H {\scriptsize II} region. We refer to the catalog of \citet{2014ApJS..212....1A} which contains $>$8,000 H {\scriptsize II} regions and H {\scriptsize II} region candidates. The catalog was made based on Wide-Field Infrared Survey Explorer (WISE) observations which have sufficient sensitivity to detect mid-infrared emission from H {\scriptsize II} regions located anywhere in the Galactic disk. However, we cannot find H {\scriptsize II} regions around G034.84$-$00.95 in Figures \ref{fig:6} and \ref{fig:5}. \citet{2015ApJS..221...26A} listed four OSC H {\scriptsize II} regions in the Galactic longitude range 30$^{\circ}$ $\leqq l \leqq$ 42$^{\circ}$. Only one H {\scriptsize II} region in G039.183$-$01.422 shows negative Galactic latitude among the four H {\scriptsize II} regions.

We also checked large scale distribution of H {\scriptsize I} emission in Galactic coordinates, since \citet{2022PASJ...74..209S} found a H {\scriptsize I} stream toward the direction of noncircular motion of G050.28$-$00.39 (see Fig. 6 of \citealp{2022PASJ...74..209S}). We could not find any H {\scriptsize I} stream below G034.84$-$00.95 and the Galactic plane. 

We failed to find a robust origin of the fast vertical motion of G034.84$-$00.95 at this moment. Further astrometric observations toward the OSC arm are required in the future. For instance, G039.183$-$01.422 contains both an H {\scriptsize II} region and an H$_{2}$O maser, and is a valuable candidate for VLBI astrometry (see Figures \ref{fig:6} and \ref{fig:5}).

\section{Summary}
We conducted VLBI astrometry toward the star-forming region G034.84$-$00.95 (Table \ref{table:4} 
) for studying the structure and kinematics of the Extreme Outer Galaxy. We overcame observational difficulty reported in previous VERA astrometry for a source close to the celestial equator \citep{2011PASJ...63..513K} by using KaVA consisting of VERA and KVN (see Fig. \ref{fig:7}). We measured the proper motion of the source to be ($\mu_{\alpha} \rm{cos}\delta$, $\mu_{\delta}$) = ($-$1.61$\pm$0.18, $-$4.29$\pm$0.16) mas yr$^{-1}$ in equatorial coordinates (Table \ref{table:2} and Fig. \ref{fig:1}), whereas we could not obtain the reliable parallax (0.040$^{+0.061}_{-0.028}$ mas). 

We estimated the source's kinematic distances through LSR velocity $V_{\rm{LSR}}$ and proper motion in the direction of Galactic longitude $\mu_{l}$, which are $d_{V_{\rm{LSR}}}$ = 17.7$^{+2.0}_{-1.6}$ kpc and $d_{\mu_{l}}$ = 18.9$\pm$1.1 kpc, respectively (Table \ref{table:2} and Fig. \ref{fig:1}). We determined the 2D kinematic distance of the source to be 18.6$\pm$1.1 kpc by the variance weighted average of both $d_{V_{\rm{LSR}}}$ and $d_{\mu_{l}}$ (Fig. \ref{fig:2}). The 2D kinematic distance places the source in the Outer-Scutum-Centaurus arm. 

We evaluated the assumption that G034.84$-$00.95 is a nearby source ($d \sim$ 2.5 kpc) and is physically related to G034.79$-$01.38 ($d$ = 2.6$\pm$0.1 kpc), G035.19$-$00.74 ($d$ = 2.2$\pm$0.2), and the supernova remnant W44 ($d \sim$ 3.1 kpc) (Table \ref{table:5} and Fig. \ref{fig:3} 
). This is because the source shows negative latitude although the outer Galactic disk is warped toward the north Galactic pole in the 1st Galactic quadrant. G034.84$-$00.95 should show large ($>$80 km s$^{-1}$) noncircular motion parallel to the disk if the source is nearby. On the other hand, G034.79$-$01.38 ($d$ = 2.6$\pm$0.1 kpc) and G035.19$-$00.74 ($d$ = 2.2$\pm$0.2) display smaller noncircular motion $<$ 10 km s$^{-1}$. Backwinding the noncircular motion of G034.84$-$00.95 does not intersect the SNR W44 in Fig. \ref{fig:3}. Thus, we rejected the case for a nearby distance ($d \sim$ 2.5 kpc) for G034.84$-$00.95. Our conclusion is supported by the fact that \citet{2015ApJS..221...26A} found a H {\scriptsize II} region associated with the OSC arm at a negative Galactic latitude (i.e., G039.183$-$01.422).

We discussed possible origins of the fast vertical motion in G034.84$-$00.95 ($W_{s}$ = $-$38$\pm$16 km s$^{-1}$ at $d_{\rm{2D}}$ = 18.6$\pm$1.1 kpc), which are the expansion of the superbubble (i.e., supernova explosions) and cloud collision with the Galactic disk. Gas circulation may occur between the outer Galactic disk around G034.84$-$00.95 with a Galactocentric distance of 12.8 kpc and halo by considering the fact that the H {\scriptsize I} supershell GS033+06$-$49 is located at a kinematic distance similar to that of G034.84$-$00.95. However, we could not find a physical origin of the vertical motion for the following reasons. We could not find an H {\scriptsize II} region around G034.84$-$00.95 (Figures \ref{fig:6} and \ref{fig:5}). Also, we could not confirm H {\scriptsize I} streams toward the direction of the noncircular motion of G034.84$-$00.95. Thus, further VLBI astrometry toward the OSC arm is important in the future. Another target, G039.183$-$01.422, associated with the OSC arm, is a valuable candidate for this purpose since the source contains H {\scriptsize II} region and H$_{2}$O (22 GHz) masers.

\begin{ack}
  We deeply thank the anonymous referee for thoughtful and constructive comments which greatly improved the quality and robustness of the paper. We acknowledge Dr. Yoshiaki Tamura for analyzing JMA data for tropospheric calibration of KVN. We deeply thank Dr. Mark J. Reid, Dr. Tomoya Hirota and Dr. Koichiro Sugiyama for stimulating discussions and valuable comments. We acknowledge EAVN (East Asian VLBI Network) project members for their support on observations, correlations, and data reductions. Data analysis was in part carried out on a common use data analysis computer system at the Astronomy Data Center, ADC, of NAOJ. 
This research made use of sedcreator \citep{2022arXiv220511422F}.
  
  BZ was supported by the National Natural Science Foundation of China (Grant No. U2031212 and U1831136),
and Shanghai Astronomical Observatory, Chinese Academy of Sciences (Grant No. N2020-06-19-005).

\end{ack}


\bibliographystyle{apj} 
\bibliography{reference}




\appendix
\section{Supplemental materials}
\label{appendix:1}

\begin{figure}[tbhp] 
 \begin{center} 
     \includegraphics[scale=0.95]{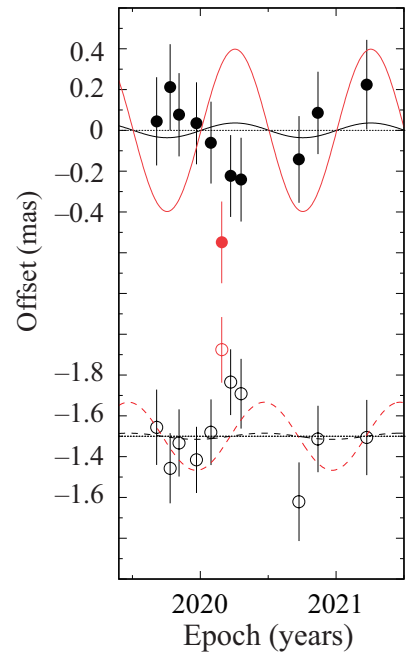} 
\end{center} 
\caption{ Same as Fig. \ref{fig:1} (middle), but with a parallax amplitude of 0.4 mas superimposed (red solid and dashed curves). Black solid and dashed curves display a parallax amplitude of 0.04 mas. Based on reduced chi-square values, a model with the smaller parallax amplitude can better explain the observational results (see the text for details).}
\label{fig:9} 
\end{figure}

\begin{figure*}[tbhp] 
 \begin{center} 
     \includegraphics[scale=1.0]{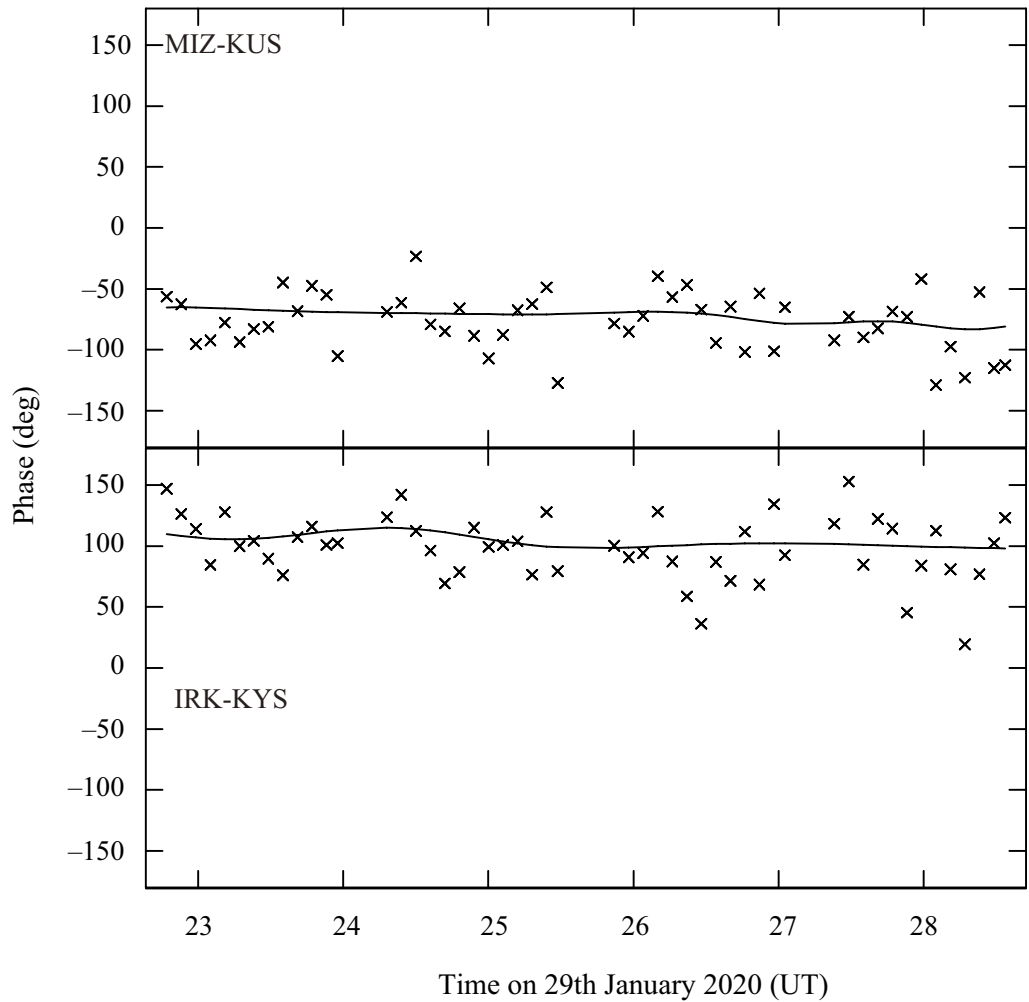} 
\end{center} 
\caption{ Corrected phases by phase referencing are shown for G034.84$-$00.95 where horizontal axis is time in (UT). {\bf(Top)} and {\bf(Bottom)} panels display the phases for Mizusawa-Ulsan and Iriki-Yonsei baselines, respectively. The data of epoch E (see Table \ref{table:4}) is used. Each cross showing the phase, is created by integrating visibilities over 6 (min) while black curves show models taken from AIPS CC (Clean Component) files.}
\label{fig:11} 
\end{figure*}

We show supplemental materials to further document observational results of G034.84$-$00.95 (Table \ref{table:6}; Figures \ref{fig:11} and  \ref{fig:9}).


\begin{table}[tbp]
\caption{Astrometric results of G034.84$-$00.95 relative to J1851+0035.}
\label{table:6}
\begin{tabular}{cccc}
\hline
\hline
DOY &\multicolumn{2}{c}{Position offset} &$V_{\rm{LSR}}$  \\ 
\cline{2-3}
       &$\Delta \alpha$&$\Delta \delta$ & \\
(day)&(mas)&(mas) &km s$^{-1}$       \\
\hline
    248     &0.682$\pm$0.082&0.178$\pm$0.092 &$-$44.0                   \\
    284     &0.689$\pm$0.068          &$-$0.446$\pm$0.063                     &$-$44.0                   \\
    308     &0.448$\pm$0.040          &$-$0.603$\pm$0.039                     &$-$44.0                   \\
    355     &0.197$\pm$0.020          &$-$1.238$\pm$0.027                     &$-$44.0                   \\
    394     &$-$0.071$\pm$0.015       &$-$1.560$\pm$0.017                &$-$44.0                   \\
    423     &$-$0.656$\pm$0.014       &$-$1.502$\pm$0.018                &$-$44.0 \\  
    447     &$-$0.470$\pm$0.018       &$-$1.936$\pm$0.020                &$-$44.0                   \\
    475     &$-$0.612$\pm$0.047         &$-$2.321$\pm$0.060                &$-$44.0                   \\
    631     &$-$1.206$\pm$0.072       &$-$4.681$\pm$0.108                &$-$44.0                   \\
    682     &$-$1.204$\pm$0.026       &$-$4.972$\pm$0.029                &$-$44.0                   \\
    813     &$-$1.648$\pm$0.091       &$-$6.501$\pm$0.092                &$-$44.0                   \\

\hline

\multicolumn{4}{@{}l@{}}{\hbox to 0pt{\parbox{80mm}{\footnotesize
\par\noindent \\
Column 1: day of year since 2019; Columns 2-3: position offset values of G034.84$-$00.95 in R.A. and Decl., respectively, relative to ($\alpha$, $\delta$) = (\timeform{18h58m17.6726s}, \timeform{01D16'06.400''}) in J2000. Positional errors are statistical errors; Column 4: LSR velocity ($V_{\rm{LSR}}$) of the maser emission.\\
}\hss}}

\end{tabular}
\end{table}

\onecolumn
\section{A Bayesian analysis}
\label{appendix:2}
We determined up to seven model parameters (i.e., parallax, reference positions, proper motion components, and systematic errors in R.A. and Decl., respectively) with Bayesian statistics (see a part of the model parameters in Table \ref{table:7}). For this purpose, we referred to a Fortran program developed by the VLBA BeSSeL project (Dr. Mark J. Reid). Note that we developed a Python program for our analysis. We used the Python package \texttt{lmfit} (\citealp{2016ascl.soft06014N}) for fitting models to R.A. and Decl. data simultaneously. Based on Bayes's theorem, the posterior probability distribution is described as

\begin{equation}
p(\bm{\theta}|\lbrace y_{i}\rbrace^{N}_{i=1}) = \frac{p(\lbrace y_{i}\rbrace^{N}_{i=1}|\bm{\theta})p(\bm{\theta})}{p(\lbrace y_{i}\rbrace^{N}_{i=1})} \sim p(\lbrace y_{i}\rbrace^{N}_{i=1}|\bm{\theta})p(\bm{\theta}) \nonumber 
\end{equation}
where $p(\lbrace y_{i}\rbrace^{N}_{i=1}|\bm{\theta})$ and $p(\bm{\theta})$ are the likelihood function and prior, respectively. Note that $p(\lbrace y_{i}\rbrace^{N}_{i=1})$ is a normalization constant. The model parameters are contained in $\bm{\theta}$, and $y_{i}$ show observational results. \\

To identify outlier data, we firstly assumed that data uncertainties obeyed a probability density function (PDF) with Lorentzian-like wings, which makes the fits insensitive to outliers (see ``conservative formulation'' of \citealp{2006book...58..Sivia}). The observational results in epoch ``F'' (see Table \ref{table:4}) deviated by $>$3-$\sigma$ from the PDF. We made two data sets with and without the outlier, and evaluated both the data sets in Table \ref{table:7}.\\

To draw posterior probability distributions for individual model parameters, we used the python package \texttt{emcee} which is an MIT licensed pure-Python implementation of Goodman $\&$ Weare's Affine Invariant Markov chain Monte Carlo (MCMC) Ensemble sampler (\citealp{2013PASP..125..306F}).\\

Since the posterior consists of the likelihood and the prior(s), we explain our definitions of the likelihood and the prior(s) in the following subsections. 

\subsection{Likelihood function, $p(\lbrace y_{i}\rbrace^{N}_{i=1}|\bm{\theta})$}
We used a Gaussian likelihood as

\begin{equation}
p(\lbrace y_{i}\rbrace^{N}_{i=1}|\bm{\theta}) = \frac{1}{\sqrt{2\pi(\sigma^{2}_{i}+\sigma^{2}_{{\rm sys}})}}{\rm exp}\left[-\frac{(\lbrace y_{i}\rbrace^{N}_{i=1}-\hat{y}_{i})^{2}}{2(\sigma^{2}_{i}+\sigma^{2}_{{\rm sys}})}\right] 
\end{equation}

where $\hat{y}_{i}$ and $\sigma_{{\rm sys}}$ are model (e.g., see equations 1-2 in \citealp{2012PASJ...64....7K}) and systematic error, respectively, for R.A. or Decl. data. The total likelihood is the product of all the individual likelihoods. For computing this, we show the natural logarithm of the total likelihood as

\begin{equation}
{\rm ln}\prod_{i=1}^{N}p(\lbrace y_{i}\rbrace^{N}_{i=1}|\bm{\theta}) = -\frac{1}{2}\sum_{i=1}^{N}\left[\frac{(\lbrace y_{i}\rbrace^{N}_{i=1}-\hat{y}_{i})^{2}}{\sigma^{2}_{i}+\sigma^{2}_{{\rm sys}}}+{\rm ln}(2\pi\lbrace\sigma^{2}_{i}+\sigma^{2}_{{\rm sys}}\rbrace)\right].
\end{equation}

\subsection{Prior probability distributions, $p(\bm{\theta})$}
Priors reflect our knowledge about the model parameters. We used a flat prior for reference positions in R.A. and Decl., respectively. For systematic errors in R.A. or Decl., we adopted a Jeffreys prior as

\begin{equation}
p(\sigma_{{\rm sys}}) \propto \frac{1}{\sigma_{{\rm sys}}}.
\end{equation}

\citet{2015PASP..127..994B} recommended not to use a flat prior for the distance converted via the reciprocal of the trigonometric parallax (i.e., $d$ = $\frac{1}{\pi}$). This is because the posterior probability distribution of the distance based on a flat prior shows a large tail toward large distance if the fractional parallax error is greater than 20$\%$ (see Fig. 1 of \citealp{2015PASP..127..994B}). Thus, we assumed the following prior:
\begin{equation}
  \label{Eq:1}
p(\pi)
\left\{ 
\begin{array}{ll}
1 & (0.01 \leq \pi) \\
\left(\frac{\pi}{0.01}\right)^{2} & (0.01*0.1 \leq \pi < 0.01) \\
\left(\frac{0.01*0.1}{0.01}\right)^{2} & (\pi < 0.01*0.1)
\end{array} \right.
\end{equation}
for the parallax. Note that the parallax of 0.01 mas converts to a distance of 100 kpc. The distance is well larger than the size of the Milky Way, and perhaps we could use a larger threshold than 0.01 mas. This prior naturally forces the parallax to be positive. We evaluated both a flat prior and Eq. \ref{Eq:1} for the parallax in Table \ref{table:7}.\\

Regarding proper motion components in R.A. and Decl. (i.e., slopes in time vs. positional offset values), we adopted the following prior as 
\begin{equation}
p(\mu) \propto (1+\mu)^{-3/2}.
\end{equation}
This prior allows a proper motion not to be biased to a larger value as discussed in an unpublished paper, Jaynes, E. T., (1991)\footnote[8]{The unpublished paper (``Straight Line Fitting - A Bayesian Solution'') can be downloaded from the following URL maintained by Dept. Of Chemistry and Radiology, Washington University:\url{https://bayes.wustl.edu/etj/node2.html}}. Note that the proper motion (i.e., slope in a line fitting) is not biased even with a flat prior if the number of data is enough.

\begin{table*}[htbp] 
\caption{Parallax and proper motion fits with different Bayesian approaches. \hspace{10em}} 
\begin{center} 
\label{table:7} 
\small 
\begin{tabular}{lccccccccccc} 
\hline 
\hline

 ID                 &Data &$\pi$ & $\mu_{\alpha} \rm{cos}\delta$       &$\mu_{\delta}$&\multicolumn{2}{c}{Systematic errors} &$\chi^{2}_{\nu}$&AIC&BIC&$\pi$ &Memo            \\
\cline{6-7}         & &&    &&R.A.&Decl.&&&&prior &\\
        & &(mas)&(mas yr$^{-1}$)    &(mas yr$^{-1}$)&(mas) &(mas)                                          \\
 \hline
 \multicolumn{6}{l}{\bf{ALL observational epochs are used.}}		\\
A1n                                   &RA                                    &$-$0.16$\pm$0.08  &$-$1.49$^{+0.15}_{-0.14}$  &$-$ &0.20&$-$&1.1&1.4&3.0&no                     \\
A1y    &RA   &0.03$^{+0.06}_{-0.15}$   &$-$1.54$\pm$0.18&$-$ &0.25&$-$&1.1&6.8&8.4&yes                     \\
A2n                                   &Decl                                  &0.28$^{+0.36}_{-0.37}$    &$-$        &$-$4.32$\pm$0.15&$-$&0.22&1.1&3.0&4.6&no     \\
A2y   &Decl &0.38$^{+0.32}_{-0.24}$    &$-$        &$-$4.33$\pm$0.15&$-$&0.21&1.2&2.9&4.5&yes     \\ 
A3n                                   &RA$\&$Decl       &$-$0.13$^{+0.09}_{-0.08}$        &$-$1.50$\pm$0.15        &$-$4.30$\pm$0.16&0.20&0.23&1.1&4.1&11.7&no     \\
A3y   &RA$\&$Decl      &0.04$^{+0.06}_{-0.03}$        &$-$1.56$^{+0.18}_{-0.19}$  &$-$4.31$\pm$0.15&0.26&0.21&1.1&8.6&16.3&yes     \\ \\

 \multicolumn{12}{l}{\bf{Observation epoch ``F'' (see Table \ref{table:4}) is removed with conservative formulation of \citep{2006book...58..Sivia}.}}		\\
 F1n     &RA                                     &$-$0.10$^{+0.07}_{-0.06}$        &$-$1.55$^{+0.12}_{-0.11}$              &$-$&0.14&$-$&1.3&$-$3.0&$-$2.0&no                     \\
 F1y    &RA   &0.03$^{+0.05}_{-0.02}$   &$-$1.60$\pm$0.14&$-$ &0.19&$-$&1.2&1.1&2.3&yes                  \\
 F2n                                   &Decl                                &0.33$^{+0.27}_{-0.28}$        &$-$     &$-$4.31$\pm$0.12&$-$&0.16&1.2&$-$2.0&$-$1.0&no                     \\
 F2y    &Decl &0.38$^{+0.25}_{-0.21}$ &$-$    &$-$4.31$\pm$0.11&$-$&0.15&1.3&$-$2.3&$-$1.1&yes                     \\
  F3n &RA$\&$Decl &$-$0.08$^{+0.07}_{-0.06}$&$-$1.56$^{+0.12}_{-0.11}$ &$-$4.29$\pm$0.13&0.15&0.18&1.2&$-$4.7&2.3&no&best     \\
F3y &RA$\&$Decl &0.04$^{+0.06}_{-0.03}$ &$-$1.61$\pm$0.14 &$-$4.29$\pm$0.12&0.20&0.16&1.1&$-$1.1&5.9&yes&preferred                     \\

\hline 
\multicolumn{4}{@{}l@{}}{\hbox to 0pt{\parbox{165mm}{\normalsize
\par\noindent
\\
Column 1: fitting ID; Column 2: data type (RA = only R.A. data is used; Decl = only Decl. data is used; RA$\&$Decl = both R.A. and Decl. data are used); Column 3: parallax; Columns 4-5; proper motion components in R.A. and Decl., respectively; Column 6-7: systematic errors in R.A. and Decl. respectively; Column 8: the reduced chi-square value; Column 9: Akaike's Information Criterion; Column 10: Bayesian Information Criterion; Column 11: parallax prior (no = uniform prior; yes = non-uniform prior, see the text for details); Column 12: memo. \\
}\hss}}
\end{tabular} 
\end{center} 
\end{table*} 



\subsection{Results and Summary}
\begin{figure*}[tbhp] 
 \begin{center} 
     \includegraphics[scale=1.0]{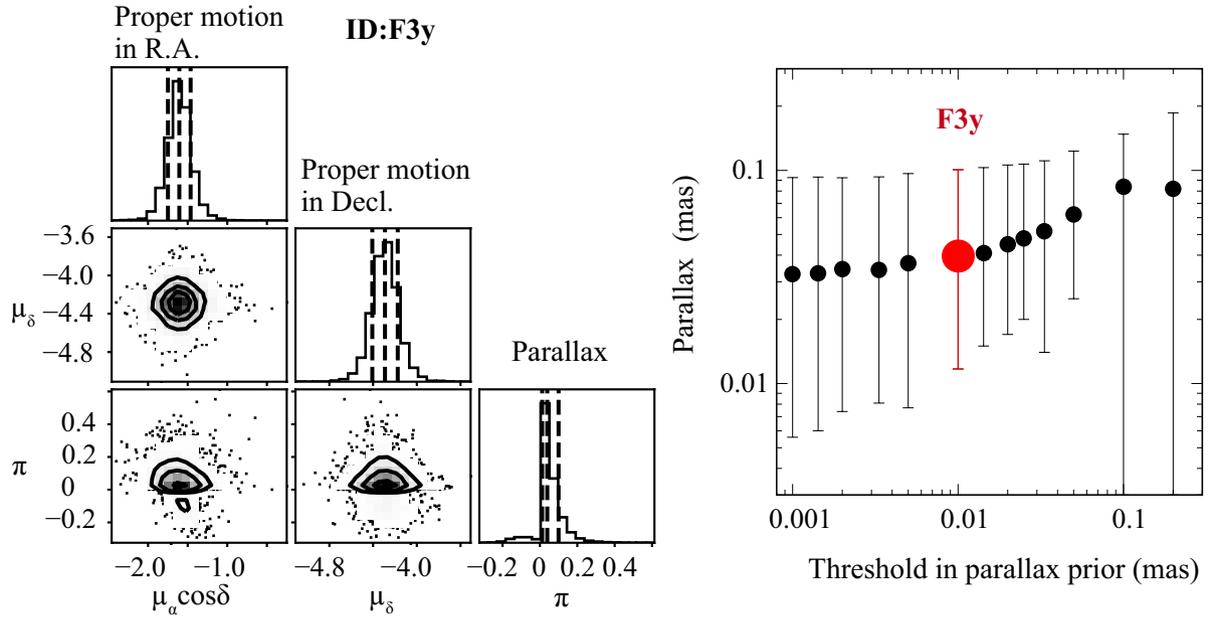} 
\end{center} 
\caption{ {\bf(Left)} Posterior probability distributions for proper motion components in R.A. and Decl., and parallax in the case of fit ID ``F3Y'' (see Table \ref{table:7}). Vertical dashed lines show the 16th, 50th, and 84th percentiles of the samples for individual model parameters. {\bf(Right)} Parallax results are shown as a function of the threshold value in the parallax prior (i.e., eq. \ref{Eq:1}). Red circle displays the result in the case of fit ID F3Y where the parallax threshold is 0.01 mas.}
\label{fig:8} 
\end{figure*}

Based on the likelihood and priors explained above, we draw posterior probability distributions of individual model parameters in the twelve Bayesian approaches (see Table \ref{table:7} and Fig. \ref{fig:8}). Proper motion results are consistent to each other in the twelve approaches while we cannot obtain a reliable parallax result in any approaches. \\

The fit ID ``F3n'' (Table \ref{table:7}) shows the best result based on AIC (Akaike's Information Criterion) and BIC (Bayesian Information Criterion) values. However, the parallax result, $-$0.08$^{+0.07}_{-0.06}$ mas, is physically implausible in the case of fit ID F3n. Thus, we adopt the fit ID ``F3y'' where the parallax result is 0.04$^{+0.06}_{-0.03}$ mas. Note that the proper motion error of R.A. is larger in the case of F3y compared to F3n. We varied the threshold value of the parallax prior (i.e., eq. \ref{Eq:1}) between 0.001 mas and 0.2 mas, and confirmed that our parallax result in the ID F3y was well converged (see the right-hand side of Fig. \ref{fig:8}).

\section{Effect of maser's internal motion on the 2D kinematic distance of G034.84$-$00.95}
\label{appendix:3}

\begin{table*}[htbp] 
\caption{
2D kinematic distances for G034.84$-$00.95.} \hspace{10em}
\begin{center} 
\label{table:6b} 
\small 
\begin{tabular}{cccccc} 
\hline 
\hline 
 $V_{\rm{LSR}}$      &$\mu_{l}$ & \multicolumn{3}{c}{Kinematic distance}      &Memo \\ \cline{3-5}
   
                &         &$d_{V_{\rm{LSR}}}$ &$d_{\mu_{l}}$ &$d_{\rm{2D}}$ \\
(km s$^{-1}$)    &(mas yr$^{-1}$)    & (kpc)&(kpc)&(kpc)                                          \\
  \hline
$-$45.1$\pm$13.2 &$-$4.55$\pm$0.25 &17.7$^{+2.0}_{-1.6}$&18.9$\pm$1.1&18.6$\pm$0.8\footnotemark[*]&Our best estimate\footnotemark[\dag]. \\

$-$5.1$\pm$13.2 &$-$4.55$\pm$0.25 &13.8$\pm$1.0&18.9$\pm$1.1&16.1$\pm$3.6\footnotemark[*]&$V_{\rm{LSR}}$ is shifted by 40 km s$^{-1}$. \\

$-$45.1$\pm$13.2 &$-$4.10$\pm$0.25 &17.7$^{+2.0}_{-1.6}$&21.1$\pm$1.3&20.1$\pm$2.4\footnotemark[*]&$\mu_{l}$ is shifted by 40 km s$^{-1}$ (= 0.45 mas yr$^{-1}$ at $d$ = 18.9 kpc). \\

\hline 
\multicolumn{6}{@{}l@{}}{\hbox to 0pt{\parbox{165mm}{\normalsize
\par\noindent
\\
Column 1: LSR velocity; Column 2; proper motion in the direction of Galactic longitude; Columns 3-4: kinematic distances through a LSR velocity and proper motion in the direction of Galactic longitude, respectively (see the text for details); Column 5: the weighted average of both Columns 3 and 4; Column 6: Memo.  \\ 
\footnotemark[*]  To examine scatters between both kinematic distances (i.e., $d_{V_{\rm{LSR}}}$ and $d_{\mu_{l}}$), each error represents the standard deviation.  \\
\footnotemark[\dag] An average noncircular motion of a Galactic spiral arm (i.e., 13 km s$^{-1}$; \citealp{2019ApJ...876...30S}) is added in quadrature to statistical errors of $V_{\rm{LSR}}$ and $\mu_{l}$ (see the text for details). \\

}\hss}}
\end{tabular} 
\end{center} 
\end{table*}

To estimate 2D kinematic distance of G034.84$-$00.95, we considered not only the systematic error of the motion of central star (i.e., 10 km s$^{-1}$), but also the noncircular motion of a Galactic spiral arm (i.e., 13 km s$^{-1}$) in the section \ref{Results}. However, this estimation may be optimistic because 22-GHz H$_{2}$O masers are typically generated in outflows of tens of km s$^{-1}$. Such high-velocity maser features are defined as $|V$ $-$ $V_{\rm{LSR}}|$ $\geqq$ 30 km s$^{-1}$ (e.g., \citealp{2010MNRAS.407.2599C}; \citealp{2011MNRAS.418.1689U}).

To more conservatively examine the effect of the maser's internal motion on the estimate of 2D distance ($d_{2D}$), we shifted LSR velocity $V_{\rm{LSR}}$ and the proper motion in the direction of Galactic longitude $\mu_{l}$ individually by 40 km s$^{-1}$ in Table \ref{table:6b}. It shows that standard deviations of 2D kinematic distances are increased (16.1$\pm$3.6 kpc or 20.1$\pm$2.4 kpc) relative to our estimate (i.e., 18.6$\pm$0.8 kpc) if only $V_{\rm{LSR}}$ or $\mu_{l}$ is shifted by 40 km s$^{-1}$. All the 2D kinematic distances are consistent with each other within errors.  

The small standard deviation of our distance estimate suggests that our estimate is less affected by the maser's internal motion. Of course, we cannot rule out the possibility that the effect of the maser's internal motion appears in the same manner in both directions of the line of sight and Galactic longitude. To confirm the robustness of our distance estimate for G034.84$-$00.95, the trigonometric parallax measurement of the source will be desired as was previously done for G007.47+00.06 where $d_{2D}$ is 20$\pm$2 kpc \citep{2016PASJ...68...60Y} and the parallax distance $d_{\pi}$ is 20.4$^{+2.8}_{-2.2}$ kpc \citep{2017Sci...358..227S}.

\section{SED fits of G034.84$-$00.95 at different distances}
\label{appendix:4}

\begin{figure*}[tbhp] 
 \begin{center} 
     \includegraphics[scale=1.0]{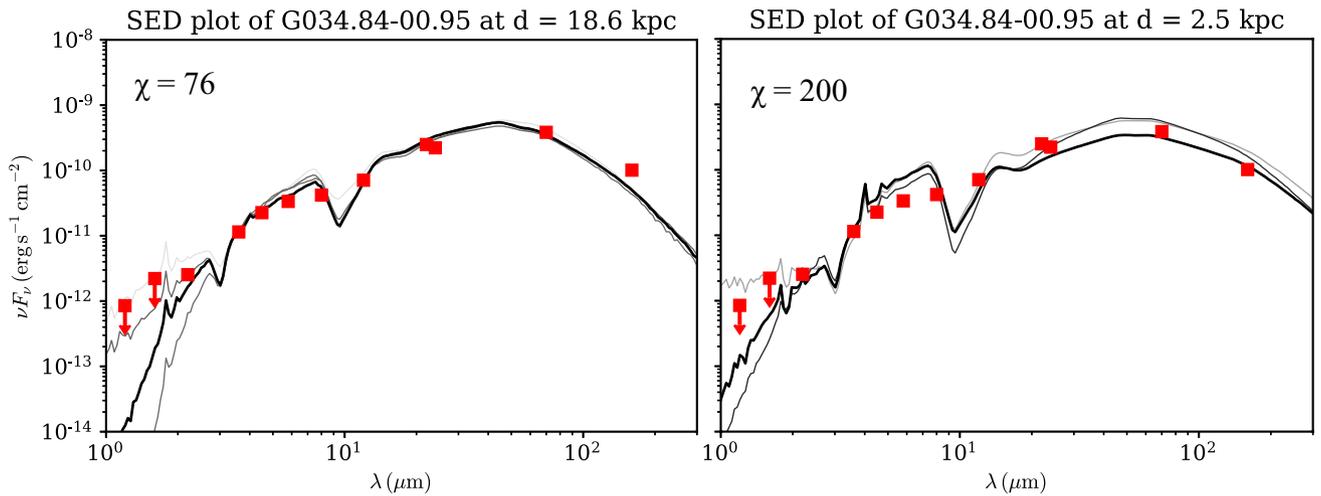} 
\end{center} 
\caption{ Results of SED fits are shown for G034.84$-$00.95 at different distances 18.6 kpc {\bf(Left)} and 2.5 kpc {\bf(Right)}. Observational results (red squares) and models (curves) are taken from Tables \ref{table:8} and \ref{table:9}, respectively. The resultant chi-square value is shown on the upper left corner of each plot. For the SED fits, we used the Python package \texttt{sedcreator} \citep{2022arXiv220511422F}. }
\label{fig:14} 
\end{figure*}

To evaluate the reliability of bolometric luminosity estimates in Table \ref{table:5}, we independently determine bolometric luminosity with a Spectral Energy Distribution fit. We used the Python package \texttt{sedcreator} \citep{2022arXiv220511422F} and an extinction law by \citep{1994ApJ...422..164K} for the SED fits. Photometric results of G034.84$-$00.95 are summarized in Table \ref{table:8}, while the results of SED fits are shown in Fig. \ref{fig:14} and Table \ref{table:9}.

The results of SED fits guarantee that our original estimates for the bolometric luminosity of G034.84$-$00.95 (i.e., Table \ref{table:5}) are unchanged. A bolometric luminosity $L_{bol}$ $>$10$^{4}$$L_{\odot}$ is obtained at a distance of 18.6 kpc, while $L_{bol}$ $\sim$10$^{3}$$L_{\odot}$ is estimated at a distance of 2.5 kpc. Other physical parameters obtained by the SED fits are listed in Table \ref{table:9}.

\begin{table*}[tbp]
\caption{Photometric results of G034.84$-$00.95.}
\label{table:8}
\begin{tabular}{cccccl}
\hline
\hline
$\lambda$ &$F_{\nu}$ &FWHM &Instrument (Project) &$r$&Ref.\\ 
($\mu$m)&(mJy)&(arcsec) &   &(arcsec)    \\
\hline
    1.2     &$<$ 0.34   &2 &2MASS  &0.3 &\citet{2006AJ....131.1163S}                \\
    1.6     &$<$ 1.18   &2 &2MASS   &0.3&\citet{2006AJ....131.1163S}                \\
    2.2     &1.86$\pm$0.17  &2  &2MASS  &0.3&\citet{2006AJ....131.1163S}                 \\
    3.6     &13.75$\pm$0.08          &3     &Spitzer (GLIMPSE)  &0.4 &\citet{2009yCat.2293....0S}                \\
    4.5     &33.9$\pm$1.4       &3  &Spitzer (GLIMPSE)&0.4 &\citet{2009yCat.2293....0S}                  \\
    5.8     &64.9$\pm$1.9       &3  &Spitzer (GLIMPSE)&0.4 &\citet{2009yCat.2293....0S}\\  
    8.0     &112.0$\pm$2.9       &3 &Spitzer (GLIMPSE)&0.4 &\citet{2009yCat.2293....0S}                 \\
    12.0     &283.6$\pm$4.2         &6.5        &WISE&0.3 &\citet{2012wise.rept....1C}                  \\
    22.0    &1831$\pm$27       &12              &WISE&0.3 &\citet{2012wise.rept....1C}                  \\
    24.0     &1779$\pm$32       &6                &Spitzer (MIPSGAL)&0.3 &\citet{2015AJ....149...64G}                   \\
    70.0     &8957$\pm$58       &5                &Herschel PACS &1.5 &\citet{2020yCat.8106....0H}               \\
    160.0     &5390$\pm$755       &13            &Herschel PACS &1.1 &\citet{2020yCat.8106....0H}                \\

\hline

\multicolumn{4}{@{}l@{}}{\hbox to 0pt{\parbox{175mm}{\footnotesize
\par\noindent \\
Column 1: observing wavelength; Column 2: flux density; Column 3: Full Width at Half Maximum of beam size (i.e., angular resolution); Column 4: instrument. Parenthesis indicates the name of a project; Column 5: angular separation from the position of 22 GHz water maser associated with G034.84$-$00.95; Column 6: references.\\
}\hss}}

\end{tabular}
\end{table*}

\begin{table*}[tbp]
\caption{Results of SED fits for G034.84$-$00.95.}
\label{table:9}
\begin{tabular}{cccccccccc}
\hline
\hline
ID &$\chi$ &$d$ &log$L$/$L_{\odot}$ &$A_{V}$&$M_{\rm{core}}$ &$R_{\rm{core}}$ &$m_{*}$&$\Sigma_{\rm{cl}}$ &$t_{\rm{now}}$\\ 
&&(kpc) & &(mag)&($M_{\odot}$)&(pc)&($M_{\odot}$)& g cm$^{-2}$  &($\times$10$^{5}$ years)    \\
\hline
    09\_01\_08     &76   &18.6 &4.87   &13.9    &120&0.3&24 &0.1 &5.0          \\
    08\_01\_07     &118   &18.6 &4.48   &0.8    &100&0.2&16 &0.1&4.1        \\
    11\_01\_10     &122  &18.6  &5.52  &36.5    &200&0.3&48 &0.1&6.8            \\
    06\_02\_06     &204  &18.6     &4.42 &4.1   &60&0.1&12  &0.316&1.7           \\
    08\_01\_06     &231       &18.6  &4.23 &17.8 &100&0.2&12&0.1&3.5                 \\ \\
    01\_03\_04     &200       &2.5  &3.06 &1.0   &10&0.02&4&1.0&0.7\\                   
    02\_01\_03     &375       &2.5 &2.28 &11.8   &20&0.1&2  &0.1&2.1            \\
    02\_01\_04     &398         &2.5&2.84 &15.1  &20&0.1&4   & 0.1&3.1            \\
    03\_01\_03    &477       &2.5   &2.38  &0.0  &30&0.1&2    &0.1 &1.9         \\
    04\_01\_03     &545       &2.5  &2.43 &57.0  &40&0.1&2     &0.1 &1.7           \\
    
\hline

\multicolumn{4}{@{}l@{}}{\hbox to 0pt{\parbox{160mm}{\footnotesize
\par\noindent \\
Column 1: model ID (see \citealp{2022arXiv220511422F} for detail); Column 2: the chi-square value for a model; Column 3: heliocentric distance; Column 4: bolometric luminosity where $L_{\odot}$ is the solar luminosity; Column 5: foreground extinction; Column 6: initial mass of the core where $M_{\odot}$ is the solar mass; Column 7: core radius; Column 8: protostellar mass; Column 9: surface mass density of the clump environment; Column 10: the current stellar age. \\
}\hss}}

\end{tabular}
\end{table*}

\end{document}